\documentclass{pasa}%

\usepackage{graphicx}
\usepackage{amssymb}
\usepackage{subcaption}
\usepackage{xcolor}

\title[Improved models for EoR0 foreground sources]{The MWA Long Baseline Epoch of Reionisation Survey -- I. Improved Source Catalogue for the EoR 0 field}

\author[C. R. Lynch et al.]{
C.~R.~Lynch,$^{1,2}$ T.~J.~Galvin,$^{1}$ J.~L.~B.~Line,$^{1,2}$ C.~H.~Jordan,$^{1,2}$ C.~M.~Trott,$^{1,2}$ J. K. Chege,$^{1,2}$ B.~McKinley,$^{1,2}$ M.~Johnston-Hollitt,$^{3}$ S.~J.~Tingay$^{1}$ 

\affil{$^{1}$International Centre for Radio Astronomy Research - Curtin University, 1 Turner Avenue, Bentley WA 6102, Australia}%

\affil{$^{2}$ARC Centre of Excellence for All Sky Astrophysics in 3 Dimensions (ASTRO 3D), Perth, WA 68}%

\affil{$^{3}$Curtin Institute for Computation, Curtin University, GPO Box U1987, Perth, 6845, WA, Australia}}%

\jid{PASA}
\doi{10.1017/pas.\the\year.xxx}
\jyear{\the\year}

\usepackage{aas_macros}
\usepackage{hyperref} 
\hypersetup{colorlinks,citecolor=blue,linkcolor=blue,urlcolor=blue}

\begin{document}

\begin{frontmatter}
\maketitle

\begin{abstract}

One of the principal systematic constraints on the Epoch of Reionisation (EoR) experiment is the accuracy of the foreground calibration model. Recent results have shown that highly accurate models of extended foreground sources, and including models for sources in both the primary beam and its sidelobes, are necessary for reducing foreground power. To improve the accuracy of the source models for the EoR fields observed by the Murchison Widefield Array (MWA), we conducted the MWA Long Baseline Epoch of Reionisation Survey (LoBES). This survey consists of multi-frequency observations of the main MWA EoR fields and their eight neighbouring fields using the MWA Phase II extended array.  We present the results of the first half of this survey centred on the MWA EoR0 observing field (centred at RA~(J2000)~0\textsuperscript{h}, Dec~(J2000)~$-$27$^{\circ}$). This half of the survey covers an area of 3069~degrees$^2$, with an average rms of 2.1~mJy~beam$^{-1}$. The resulting catalogue contains a total of 80824 sources, with 16 separate spectral measurements between 100 and 230~MHz, and spectral modelling for 78$\%$ of these sources. Over this region we estimate that the catalogue is 90$\%$ complete at 32~mJy, and 70$\%$ complete at 10.5~mJy. The overall normalised source counts are found to be in good agreement with previous low-frequency surveys at similar sensitivities. Testing the performance of the new source models we measure lower residual rms values for peeled sources, particularly for extended sources, in a set of MWA Phase I data. The 2-dimensional power spectrum of these data residuals also show improvement on small angular scales -- consistent with the better angular resolution of the LoBES catalogue. It is clear that the LoBES sky models improve upon the current sky model used by the Australian MWA EoR group for the EoR0 field.  
\end{abstract}

\end{frontmatter}

\section{INTRODUCTION }\label{sec:intro}
The formation of the first luminous sources and their subsequent reionisation of the intergalactic medium (IGM) is called the Epoch of Reionisation (EoR). During this time astrophysical sources became the dominant influence on the conditions of the IGM and impacted all future generations of galaxy formation and evolution \citep{Furlanetto:2006, Morales:2010}. Despite its importance, the EoR is one of the last uncharted eras in the history of the Universe.

The most promising method for investigating the EoR is through tomography of the redshifted 21 cm line of neutral hydrogen. Hydrogen is isotropic and ubiquitous, making up roughly 75$\%$ of the gas mass present in the IGM. Thus it acts as a convenient tracer of the properties of this medium \citep{Bowman:2009, Pritchard:2012}. The 21 cm line is produced via hyperfine splitting, caused by the interaction between the electron and proton magnetic moments. Fluctuations in the redshifted 21 cm signal arise from a range of different physical properties, including inhomogeneities in the gas density field, ionisation fraction, and spin temperature. These fluctuations create angular structure as well as structure in redshift space. Thus the 21 cm line traces the entire three-dimensional ionisation history of the IGM \citep{Tozzi:2000, Furlanetto:2006, Morales:2010, Pritchard:2012}. 

Detecting 21 cm emission from the EoR is a goal for many current low-frequency radio interferometers including the Low Frequency Array (LOFAR; \citet{vanHaarlem:2013}), the Murchison Widefield Array (MWA; \citet{Tingay:2013, Wayth:2018}), and the Precision Array to Probe the Epoch of Reionisation (PAPER; \citet{Parsons:2010}). Additionally, next generation radio telescopes including the Square Kilometre Array (SKA; \citet{Koopmans:2015}), and the Hydrogen Epoch of Reionisation Array (HERA; \citet{DeBoer:2017}) will have improved sensitivities over current facilities, aiming to provide not only high signal-to-noise detections of the EoR signal over multiple redshifts, but also the first three-dimensional tomographic images of the EoR. Pipeline analysis and power-spectrum upper limits from current EoR experiments using the MWA \citep{Beardsley:2016, Barry:2019, Li:2019, Trott:2020}, LOFAR \citep{Patil:2017, Gehlot:2019, Mertens:2020} and PAPER \citep{Cheng:2018, Kolopanis:2019} have highlighted several systematic challenges to making a detection. In particular, previous papers stress the requirement of highly precise sky models for calibration and foreground removal \citep{Datta:2010, Trott:2016b, Barry:2016, Patil:2016, Ewall-Wice:2017, Byrne:2019, Kern:2020}.

\begin{table*}[t!]
\caption{Details of the MWA Phase II data set used to the create new sky model for the EoR0 fields. Listed is the LoBES field number and central right ascension (RA) and declination (Dec), date of the observations, the total integration time per snapshot image, the central frequency of the observing band, and the observation IDs associated with this data.}\label{table:obs}
\begin{tabular}{ccccccc}
\hline
Field & RA & Dec     & Date  & Integration Time & Frequency 	& observation IDs\\
& (J2000) & (J2000) &  & seconds & MHz & \\
\hline
1  & 00:00:00 &$-$27:00:00 & 2017 Nov 08 2017 & 120 & 119.0/154.9  & 1194179376 -- 1194183816\\
1  & 00:00:00 &$-$27:00:00 & 2017 Nov 09 2017 & 120 & 185.6/216.3  & 1194265064 -- 1194269984\\
2  & 22:42:00 &$-$27:00:00 & 2019 Jun 24 2019 & 296 & 119.0           & 1245435024 -- 1245437096\\
2  & 22:42:00 &$-$27:00:00 & 2019 Jun 24 2019 & 296 & 154.9          & 1245437424 -- 1245439496\\
2  & 22:42:00 &$-$27:00:00 & 2017 Nov 09 2017 & 120 & 185.6/216.3  & 1194260392 -- 1194264840\\
3  & 00:00:00 &$-$07:00:00 & 2017 Nov 12 2017 & 120 & 119.0/154.9  & 1194523552 -- 1194528472\\
3  & 00:00:00 &$-$07:00:00 & 2017 Nov 13 2017 & 120 & 185.6/216.3  & 1194609720 -- 1194614640\\
4  & 01:18:00 &$-$27:00:00 & 2017 Nov 08 2017 & 120 & 119.0/154.9  & 1194184048 -- 1194188488\\
4  & 01:18:00 &$-$27:00:00 & 2017 Nov 09 2017 & 120 & 185.6/216.3  & 1194270208 -- 1194274648\\
5  & 00:00:00 &$-$47:00:00 & 2017 Nov 23 2017 & 120 & 119.0/154.9  & 1195471360 -- 1195476160\\
5  & 00:00:00 &$-$47:00:00 & 2017 Nov 25 2017 & 120 & 185.6/216.3  & 1195643688 -- 1195648608 \\
\hline
\end{tabular}
\end{table*}

For the MWA EoR experiment, the main observing fields were chosen based on their low sky temperature \citep{Bowman:2009}, and are designated EoR0 (centred at RA~(J2000)~0\textsuperscript{h}, Dec~(J2000)~$-$27$^{\circ}$) and EoR1 (centred at RA~(J2000)~4\textsuperscript{h}, Dec~(J2000)~$-$30$^{\circ}$). However, the large field of view of the MWA makes it challenging to avoid all bright extended radio galaxies and the Galactic plane entirely. Therefore the MWA EoR fields contain several bright, extended sources located near the edges of the MWA primary beam or within the primary beam sidelobes \citep{Jacobs:2016}. Because the chromaticity of interferometers becomes stronger far from the instrument pointing centre, sources located far from the primary field of view will produce more foreground contamination than sources located in the central part of the field \citep{Trott:2012, Thyagarajan:2015a, Thyagarajan:2015b, Trott:2020}. Further, \citet{Pober:2016} showed that using a foreground model that includes only sources located within the main lobe of the primary beam will be insufficient to suppress foreground power leakage and concluded that foregrounds should be considered as a wide-field contaminant. 

Additionally, detecting the 21~cm emission from the EoR requires accurate structure models for all foreground sources, but especially bright extended sources \citep{Procopio:2017, Trott:2017}. Comparing the residual power for point source models versus multi-gaussian models based on higher resolution data from TGSS ADR1 \citep{Intema:2017}, \citet{Procopio:2017} showed that mismodelling bright, extended sources in the EoR1 field as point sources contributes the majority of the foreground residual power. Most recently, \citet{Line:2020} modelled Fornax~A, one of the brightest and most complex sources located within the MWA EoR main observing fields, using shapelets and MWA data at multiple resolutions. When testing the shapelet model of Fornax~A, \citet{Line:2020}, found that the residuals in their set of real MWA test data were dominated by systematics unrelated to Fornax~A alone. However, when using a simulated MWA data set, the residual power in the EoR PS after foreground removal was reduced by two orders of magnitude at all angular scales, just by improving the Fornax~A model.

The current sky model used by the Australian MWA EoR group is derived from a set of archival multifrequency radio catalogues that were cross-matched using the Positional Update and Matching Algorithm (\textsc{puma}; \citet{Line:2017, Line:2018}). The catalogues included are 74~MHz  Very  Large  Array  Low  Frequency  Sky  Survey re-dux (VLSSr; \citet{Lane:2012}), the 843~MHz Sydney University Molonglo Sky Survey (SUMSS; \citet{Mauch:2003}), the 1.4 GHz NRAO VLA Sky Survey (NVSS; \citet{Condon:1998}), and the GLactic and Extragalactic All-sky MWA survey Extragalactic Catalogue Release 2 (GLEAM; \citet{HurleyWalker:2017}). This sky model is missing sources in the sky region bounded by declinations between 0~-- $+$30$^{\circ}$, and right ascensions between 22 -- 0~hours. This is a result of the current Australian MWA EoR sky model being based mostly on GLEAM, which similarly lacks sources within this region due to poor ionospheric conditions during observations of this region of the sky. The majority of the sky model sources are modelled using single-component models, with a subset of the sources located in the EoR1 field modelled using the multi-gaussian models from \citet{Procopio:2017}. Also included is the shapelet model for Fornax~A from \citet{Line:2020}. What is lacking from the current Australian MWA EoR sky model is accurate, high-resolution, source modelling for complex sources located within the primary beam sidelobes of the main MWA EoR fields, and wide-range spectral coverage for these more complex models. With the recent upgrade to the MWA, we can now address these issues within the current Australian MWA EoR sky model.

The Phase II upgrade to the MWA doubles the total number of antenna tiles deployed in the array to a total of 256 tiles, however the current correlator of the array can only process 128 dual-polarisation signals at a time. The array is then periodically re-configured between a compact configuration consisting of 56 MWA Phase I core tiles and 72 additional tiles arranged in two compact hexagonal cores, and an extended configuration of 72 MWA Phase I outer tiles and 56 new long baseline tiles. Further details on the upgrade to the MWA can be found in \citet{Wayth:2018} and \citet{Beardsley:2019}.

In the extended configuration the longest baseline of the MWA is 5.3~km, nearly doubling the longest baseline of Phase I. The longer baselines provide higher resolution imaging and a reduction in the classical confusion noise for the MWA. Additionally, the extended configuration does not include the original MWA Phase I core and so it no longer contains the core's many short baselines. This results in a more uniformly filled $uv$-plane as compared to the Phase I array \citep{Wayth:2018}, with an associated improvement in the synthesised beam and lower sidelobe confusion noise. 

\begin{figure}[t!]
\begin{center}
\includegraphics[width=\columnwidth]{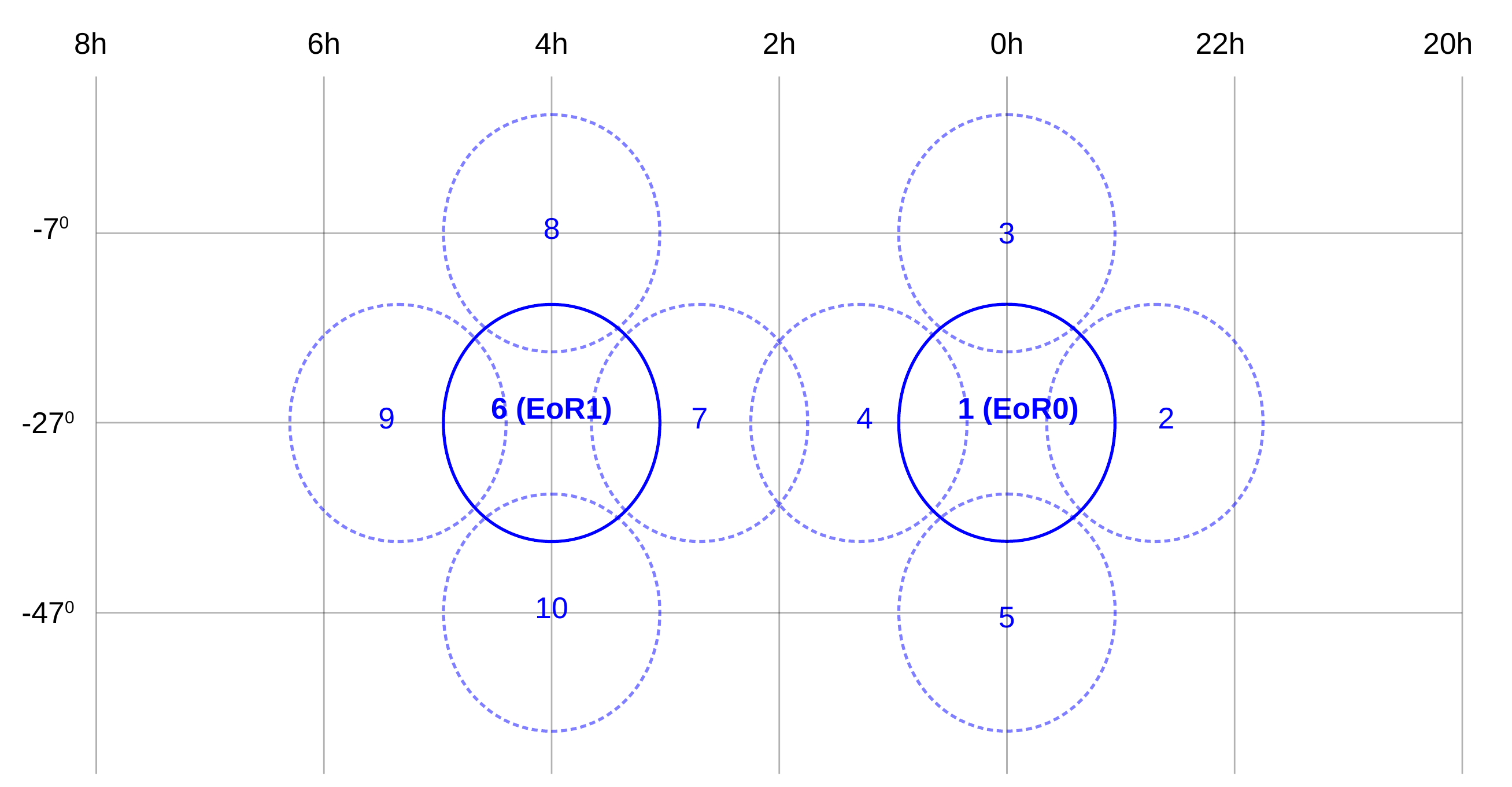}
\caption{Layout of the observing fields contained with the LoBE survey. In this paper we focus on fields 1 -- 5; these are the fields associated with the main MWA EoR observing field EoR0. Note that the MWA EoR1 observing field contains two A-Team sources, Fornax~A and Pictor~A, whose modelling and removal requires more advanced techniques than those outlined in this paper. }\label{lobes-layout}
\end{center}
\end{figure}

Taking advantage of the improved resolution and expected lower confusion noise of the longer baselines, we use the MWA Phase II extended array to conduct the Long Baselines Epoch of Reionisation survey (LoBES). This survey consists of deep multi-frequency observations of the two primary MWA EoR fields and their eight neighbouring fields (see Figure \ref{lobes-layout}). Observations of the neighbouring fields will improve our ability to calibrate and remove these contaminating sources in the main EoR field sidelobes. In this paper we present the first results from the LoBE survey for the MWA EoR0 field and its four neighbouring fields. Previous efforts to accurately model the foreground sources in the MWA EoR0 field include \citet{Offringa:2016} and \citet{Carroll:2016}. Both works used MWA Phase I data  to generate their source models. Additionally, \citet{Offringa:2016} and \citet{Carroll:2016} focused only on sources found within the main lobe of the MWA primary beam. The LoBE survey improves upon these results by using higher angular resolution imaging and includes sources over a larger area of the sky, covering the full MWA primary beam response. Results for the MWA EoR1 field will be presented in a follow-up paper (Lynch et al. \textit{in prep}).

\section{Observations and reduction}
\subsection{Observations}
We observed the MWA EoR0 field and its four flanking fields using the new extended MWA Phase II array. These fields were observed in four frequency bands covering: (1) band~1~=~103.7~--~133.1~MHz; (2) band~2~=~139.5~--~169.0~MHz; (3) band~3~=~170.2~--~199.7~MHz; (4) band~4~=~201.0~--~230.4~MHz. While the majority of these observations took place in Nov 2017, due to bad weather the two lower bands for field 2 were re-observed 24 June 2019 using Directors Discretionary Time. For all the observations we utilised the MWA's `drift and shift' observing mode, where an analog beamformer steers the main lobe of the primary beam to approximately the same sky coordinates for each pointing of the array. The sky is then allowed to drift for roughly 10~minutes, while a series of short `snapshot' observations are taken, before re-pointing. The observations for fields 1, 3 -- 5, and the upper two bands of field 2 were recorded as a set of 120~second snapshots, alternating between the two frequencies listed in Table \ref{table:obs} for each field. We observed the lower two bands for field 2 in a set of 296~second snapshots, where observations occurred consecutively, with band~1 followed by band~2. For each LoBES field, 40 minutes of observation were recorded in each frequency band. 

\subsection{Characterising the ionospheric activity}\label{sec:qametrics}

The ionosphere is a turbulent region of the Earth’s upper atmosphere, permeated by the Earth’s magnetic field and ionised by Solar radiation \citep{Kintner:1985}.  It acts as a refractive medium for incident radio waves, imparting line-of-sight refractive shifts that are proportional to the square of the wavelength of the incident radiation. Variations in ionospheric heating by the Sun lead to changes in the total electron content (TEC) of the ionosphere. The variations of the TEC towards background astronomical sources will cause angular shifts in their positions as observed by a radio interferometer \citep{Thompson:2001}. During extreme ionospheric activity, these positional offsets can be large \citep[e.g.][]{Loi:2015}. If these effects are not accounted for when combining multiple observations, the result will be to blur the resolution element for the combined image. 

Using data from the MWA, \citet{Jordan:2017} developed a quality metric to describe the degree of ionospheric activity during an observation. They found that 74$\%$ of MWA observations can be described as having little to no ionospheric activity. Similarly, \citet{Mevius:2016} used a set of LOFAR data to sample the nighttime ionospheric activity and assessed the quality of the data via the ionospheric diffractive scale, i.e. the length scale where the phase variance is 1 rad$^2$. They found that about 90$\%$ of the sample data had ionospheric diffractive scales large enough to allow for the high dynamic-range imaging required by the LOFAR EoR experiment.

To assess the ionospheric activity during the nights the LoBES data were observed, we calculated ionospheric quality metrics using the Real Time system (RTS; \citet{Mitchell:2008}) and \textsc{cthulhu} as outlined in \citet{Jordan:2017}. Removing active nights removed 12$\%$ of the total data. For the majority of the LoBES fields only a few observations were removed per band, leaving between 35~--~40~minutes of data, however for the upper two bands of the LoBES 4 field, half of the data were removed due to ionospheric activity, leaving 20~minutes per band.

\subsection{Pre-processing, Calibration, and Peeling}

We used the supercomputing facilities at the Pawsey Supercomputing Centre\footnote[1]{https://pawsey.org.au} in Perth, Western Australia to carry out data reduction and imaging on a per field, per frequency basis. The LoBES data were flagged for radio frequency interference (RFI), averaged and converted to \textsc{casa} measurement sets \citep{Mcmullin:2007}, and downloaded via the MWA All-Sky Virtual Observatory\footnote[2]{https://asvo.mwatelescope.org/} (MWA ASVO). The MWA ASVO uses the \textsc{aoflagger} algorithm \citep{Offringa:2012, Offringa:2015} to flag RFI, and perform averaging and conversion of the visibilities. The visibilities downloaded from the LoBE survey were averaged to a time resolution of 4~s and frequency resolution of 40~kHz.

We calibrated each snapshot observation using the Australian MWA EoR sky model as described in Section \ref{sec:intro}. For each observation, we generated a sky model from the cross-matched catalogue containing the brightest 200 sources for the appropriate pointing on the sky, taking into account the MWA primary beam. We generated and applied amplitude and phase solutions on a per snapshot basis using \textsc{mitchcal} \citep{Offringa:2016} which is an implementation of the direction-independent, full-polarisation algorithm described by \citet{Mitchell:2008}. After applying the calibration, we re-flagged the visibilities using \textsc{aoflagger} to flag any remaining radio frequency interference missed during the initial processing of the data. Initial calibration and imaging of the LoBES field 5 revealed additional bad data in the lowest two observing bands. Using \textsc{casa} we identified the bad data as being associated with a single tile, which was then flagged.

\subsection{Self-Calibration and Imaging}\label{sec:imaging}
We image each snapshot observation using the wide-field imager \textsc{wsclean} \citep{Offringa:2014}, which uses $w$-stacking to deal with the wide-field $w$-term effects. Recently, both multi-scale and multi-frequency deconvolution algorithms were integrated into \textsc{wsclean} \citep{Offringa:2017}. Spectral variations need to be taken into account during deconvolution due to intrinsic spectral variations of the foreground sources, the chromatic primary beam of the MWA, and the high dynamic-range required for images of the EoR foreground sky. Multiscale imaging reduces the impact of negative bowls around bright, resolved sources in our images and has better convergence properties \citep{Rich:2008}. An auto-masking algorithm is also implemented within \textsc{wsclean}, which allows for automated deep cleaning using a single run of \textsc{wsclean} and improves the multiscale cleaning by maintaining scale-dependent masks. The auto-masking algorithm first cleans down to an initial threshold set using the “auto-mask” parameter, while simultaneously recording the positions and scale of each component in a scale-dependent mask. After this first threshold is reached, the cleaning will continue down towards a final threshold set by the “auto-threshold” parameter. During this last stage of cleaning the recorded scale-dependent mask is used to constrain the cleaning \citep{Offringa:2017}. To implement these algorithms, we used the "multiscale" and "join-channels" options in \textsc{wsclean}, splitting the total 30.72~MHz bandwidth into four 7.68~MHz channels and jointly-cleaning them. This outputs four 7.68~MHz subband images as well as a full-bandwidth, MFS image. We used both these algorithms within \textsc{wsclean} to generate sets of spectral images for each snapshot observation. 

For all imaging, we used the Briggs scheme with a robust parameter of 0.0 \citep{Briggs:1995}, which provides greater image sensitivity over uniform weighting without sacrificing too much with regards to image resolution. We imaged the primary beam down to the 10$\%$ level and chose the pixel scales so that the full width at half its maximum value of the synthesised beam is sampled by at least 5 pixels. The imaging parameters for each of the four frequency bands are given in Table \ref{table:imaging}. 

\begin{table}
\caption{Imaging parameters used for each of the four frequency bands included in LoBES survey. Table columns are the central frequency, the pixel size (Cell), and the imaged field of view (FOV)}
\begin{tabular}{lcl}
\hline
Frequency & Cell           & FOV \\
 (MHz)    & (arcsec/pixel)	      & (deg$^2$)  \\
\hline
119.0 & 24.5 x 24.5 & 39.7 x 39.7 \\ 
154.9 & 18.7 x 18.7 & 37.3 x 37.3 \\
185.6 & 15.5 x 15.5 & 30.0 x 30.0 \\
216.3 & 13.3 x 13.3 & 25.0 x 25.0 \\
\hline
\end{tabular}\label{table:imaging}
\end{table}

We performed a single iteration of self-calibration for each snapshot observation. We created a shallow image, jointly-cleaning in instrumental polarisation ($XX$, $YY$, $YX$, $XY$) using an auto-mask threshold of eight times the predicted thermal noise and a final auto-threshold of five times the noise.  During this initial imaging, \textsc{wsclean} stored the best clean model in the model column of the measurement set. We then calibrated the data using \textsc{mitchcal} and the stored model. After applying these calibration solutions we performed a final round of flagging. 

We imaged the self-calibrated visibilities using the implementation of image domain gridding (IDG; \citet{Vandertol:2018}) within \textsc{wsclean}. IDG is a new gridding algorithm that makes $w$-term and $a$-term corrections computationally more efficient when using graphics processing units and allows for gridding with a time-variable beam. Additionally, \citet{Vandertol:2018} showed that IDG more accurately deconvolves sources as compared to classical gridding algorithms. For each snapshot observation, we use the IDG and the "grid-with-beam" options within \textsc{wsclean} to apply primary beam corrections using the Full Embedded Element primary beam model for the MWA \citep{Sokolowski:2017}. We image all four Stokes polarisations (IQUV) using the "link-polarisation" option to clean sources identified in Stokes I; for the remainder of the analysis we only use the outputted Stokes I and V images. These final images are cleaned down to an auto-mask threshold of four times the predicted thermal noise and a final auto-threshold of one times the noise.

\section{Mosaic creation}\label{sec:mosaic} 

Mosaic creation was performed in two steps using the software \textsc{swarp} \citep{Bertin:2002}. First, individual mosaics were created per LoBES field and frequency subband, generating a set of 16 spectral images for each LoBES field. During this initial stage of mosaicking, each snapshot image was weighted by the squared ratio of the primary beam response to the typical noise in the centre of each image. This weighting scheme accounts for the variation in the noise over the field for each snapshot image and minimises the noise in resulting mosaic \citep{Sault:1996, HurleyWalker:2017}. To create a deep image to perform source fitting, we mosaicked the eight highest frequency spectral images into a single broad-band image for each LoBES field. Before mosaicking the spectral images, we used the \textsc{convol} command in the \textsc{miriad} software suite \citep{Sault:1995} to convolve each of the spectral images to the resolution of the lowest frequency image (a resolution of 77.5~arcseconds). The combined weight maps created by \textsc{swarp} for each spectral image during the first mosaicking step were used as the image weights during this last step of mosaicking. All mosaics were formed using a slant orthographic projection. 

\subsection{Additional Corrections}
Before mosaicking the individual snapshot images, we performed additional corrections to account for residual positional offsets and flux density scale variations across the observed field of view. To perform both corrections we used \textsc{Aegean} \citep{Hancock:2012, Hancock:2018} to perform initial source finding on each snapshot image to identify unresolved sources with signal-to-noise ratio $\geq$8. We further selected only isolated sources, removing sources that have any neighbouring sources within 5 arcminutes. 

Using \textsc{puma}, we then cross-match the snapshot image catalogues with a set of radio catalogues that cover a large range in radio frequencies and included NVSS, SUMSS, GLEAM, and VLSSr. For both corrections outlined below we selected only sources identified by \textsc{puma} as having an `isolated' match type with a positional probability greater than 95$\%$ -- these are matches for which there is only one cross-match combination and are indicative of unresolved sources with no nearby confusing source at higher resolutions \citep{Line:2017}. The output from \textsc{puma} is either a FITS or VOTable that includes the spectral information collected for each base source from each matched catalogue, spectral model parameters, and the updated position of the base source based on a ranking of the matched catalogues. For our cross-matches using LoBES we chose the ranking order based on the resolution of each catalogue, where NVSS was the highest ranked catalogue due to its high angular resolution. For each image there are between 300 - 1000 sources used to perform these corrections, depending on the observation quality and frequency.

\begin{figure*}[t!]
\centering
\begin{subfigure}[b]{\textwidth}
\centering
\includegraphics[scale=0.35]{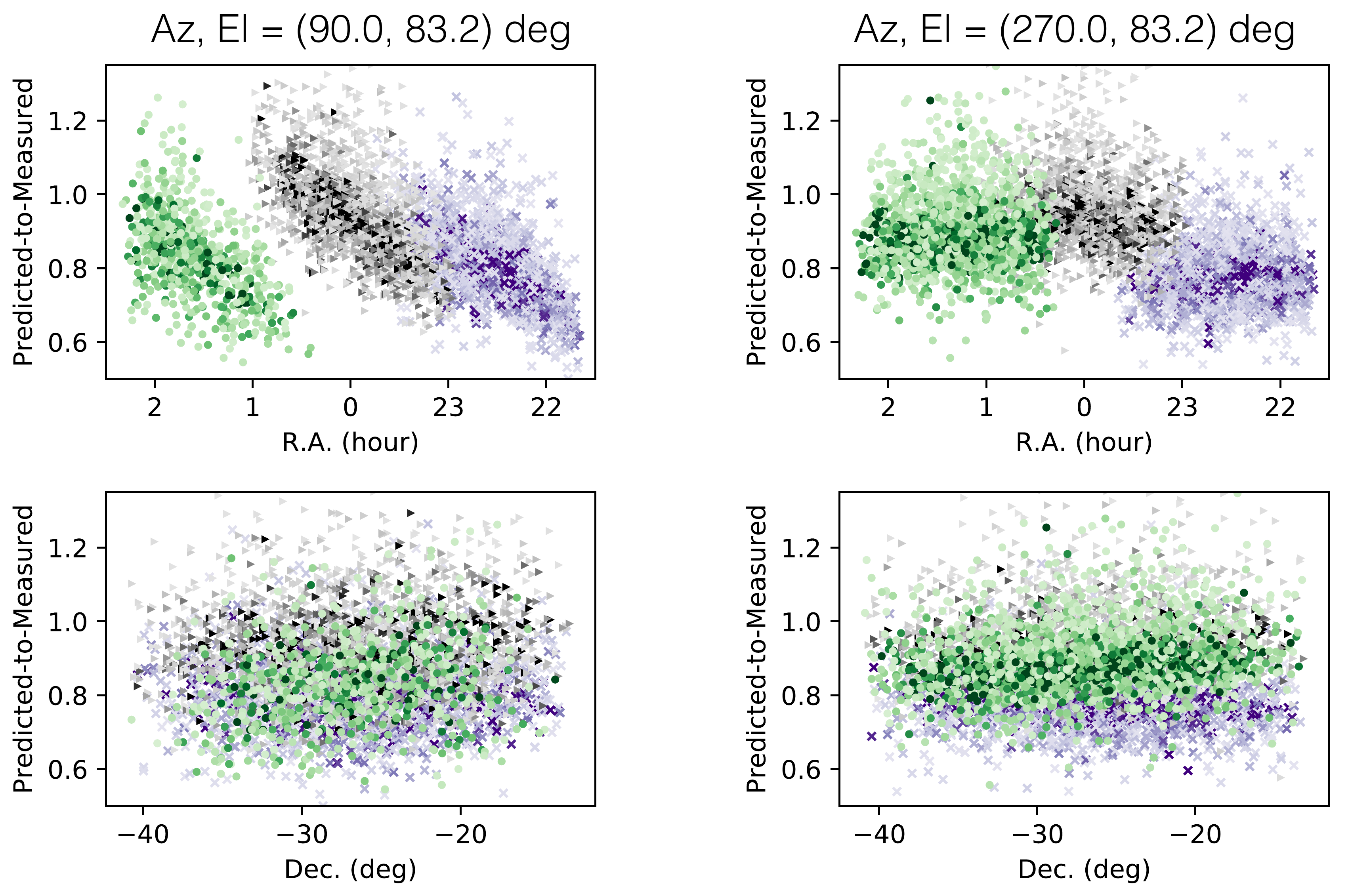}
\caption{Uncorrected}
\vspace{5mm}
\end{subfigure}

\begin{subfigure}[b]{\textwidth}
\centering
\includegraphics[scale=0.35]{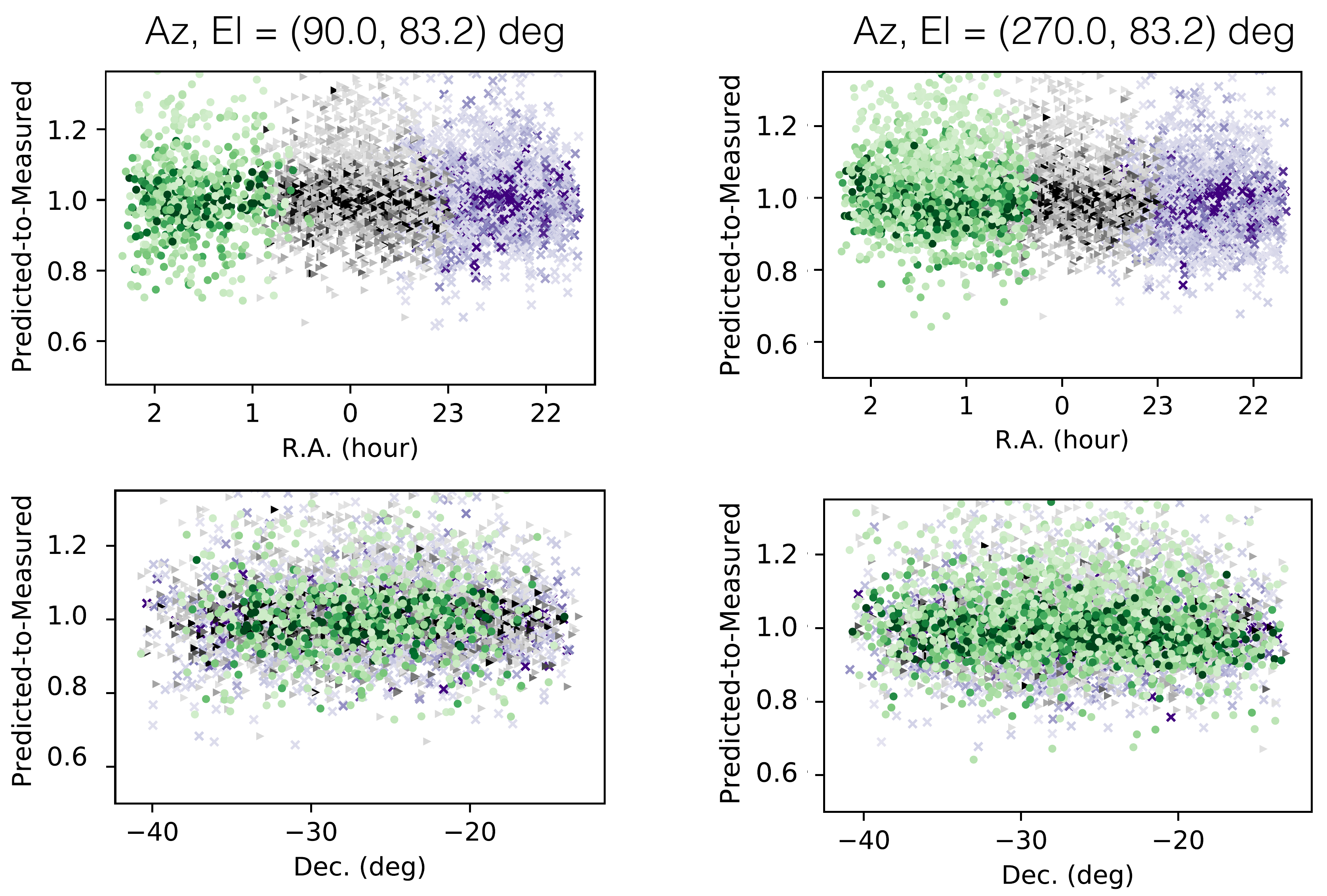}
\caption{Corrected}
\end{subfigure}
\caption{The uncorrected (a) and corrected (b) ratios of the measured LoBES flux density measurements as compared to that predicted by other multi-frequency radio surveys, as a function of right ascension (top row) and declination (bottom row). These figures are for a single 2~minute observation at 189~MHz in LoBES fields 1 (black triangles), 2 (purple x's), and 4 (green circles). The shading of the color represents the ratio of the signal-to-noise of the source, with higher ratios represented by deeper colour.} The left column and right columns are for two different MWA pointings and illustrate that the variation is also pointing dependent. \label{fig:flxvar}
\end{figure*}

\subsubsection{Astrometric}\label{sec:astro} 

Using the ionospheric quality metrics from \citet{Jordan:2017}, we have removed LoBES data associated with the most severe ionospheric conditions during the observations (see Section \ref{sec:qametrics}), however the remaining snapshot observations are still affected by the ionosphere. To correct for these ionospheric effects we do a image-based position correction using \textsc{fits\_warp} \citep{HurleyWalker:2018}. This program compares a catalogue of measured source positions from an image to a reference catalogue, generates a model of the positional offsets, and uses this model to de-distort the image. To correct our snapshot images, we feed \textsc{fits\_warp} the measured LoBES source positions and the updated position from our \textsc{puma} cross-match of the LoBES snapshot sources with NVSS, SUMSS, GLEAM and VLSSr. 

\subsubsection{Flux density scale}\label{sec:flx} 
As noted previously, \textsc{puma} reports the model parameters for a spectral model fit to the catalogue information for each matched source; the model fit to the flux densities, $S_{\nu}$, at frequencies $\nu$ is:
\begin{equation}
\ln\left(S_{\nu}\right) = \alpha \ln\left(\nu\right) + \beta
\end{equation}
where the flux densities are in Jy, the frequencies are in MHz, and $\alpha$ and $\beta$ are the spectral index and intercept, respectively, reported by \textsc{puma} along with their associated errors. Using the spectral model information reported by \textsc{puma}, we calculate the predicted flux density for each matched LoBES point source per snapshot image, at the appropriate observing frequency. Comparing the predicted flux densities to those measured in the snapshot images we find a flux density scale variation across the LoBES images. 

Figure \ref{fig:flxvar}a shows examples of the flux density scale variation at 189~MHz for LoBES fields 1, 2, and 4, in declination and right ascension, for two different MWA pointings used during the observations. We note that the flux density scale variation changes not only field to field and with frequency, but also between the pointings used for the same LoBES field (at the same frequency).  We believe that the characteristics of the observed flux density scale variation are an indication that the variation is due to residual primary beam model errors. Note that the MWA is pointed via delay steps added to the physical path lengths of individual dipoles that make up a MWA tile. This creates a series of coarse pointing adjustments when tracking a single point in the sky, each with their own primary beam shape \citep{Tingay:2013}. We expect then, that each coarse pointing will have its own associated set of primary beam errors which will be seen as not only polarisation leakage in Stokes Q, U and V but also flux density scale variations in total intensity, similar to that seen in \citet{HurleyWalker:2014, HurleyWalker:2017} and \citet{Lenc:2018}.

Note that \citet{Duchesne:2020} also found that the integrated flux densities measured using an older version of \textsc{aegean} were affected by an internal calculation of the PSF carried out and applied by \textsc{aegean}. This internal PSF correction created a position-dependent error in the integrated flux densities. We do not believe this is an issue in the analysis presented here for two reasons: (1) we used an updated version of \text{aegean} (version 2.2.3 updated 2020-07-23); (2) we performed a similar analysis of the flux density scale using a different source finder, and measured the same flux density scale variations across the images that we measured using \textsc{aegean}. The flux density scale variation we measure is independent of the source finder used. 

To model and correct for the flux density scale variation in each snapshot image we perform similar corrections as that described in \citet{Lenc:2018}, although we have updated this method to work for deconvolved sources by using the source integrated flux density rather than the peak pixel values. We first calculate the ratio between the predicted flux density from the \textsc{puma} spectral model and the LoBES measured flux density for each matched source. Using the positions of the matched LoBES sources, we then grid the flux density ratios and fit a two-dimensional quadratic surface to the grid to form a scaling map for each snapshot image. This scaling map is then applied to both the Stokes I and Stokes V image. Examples of the corrected flux density ratios are shown in Figure \ref{fig:flxvar}b for LoBES fields 1, 2, and 4 at 189~MHz -- here it is clear that our correction has removed the variation structure as a function of right ascension and declination.

\section{Generating the source catalogue}

The catalogue creation process used for the LoBE survey is similar to that used by GLEAM. First a deep, wide-band image is used to create a reference catalogue. For the LoBE survey, we created a single wide-band image, covering the frequency range 170~--~230~MHz, for each of the five fields.  These deep images minimise the image thermal noise, while still achieving high image resolution. Then the flux density for each source in this reference catalogue is measured in each of the sixteen 7.68~MHz spectral images. 

\begin{figure}[t]
\begin{center}
\includegraphics[width=\columnwidth]{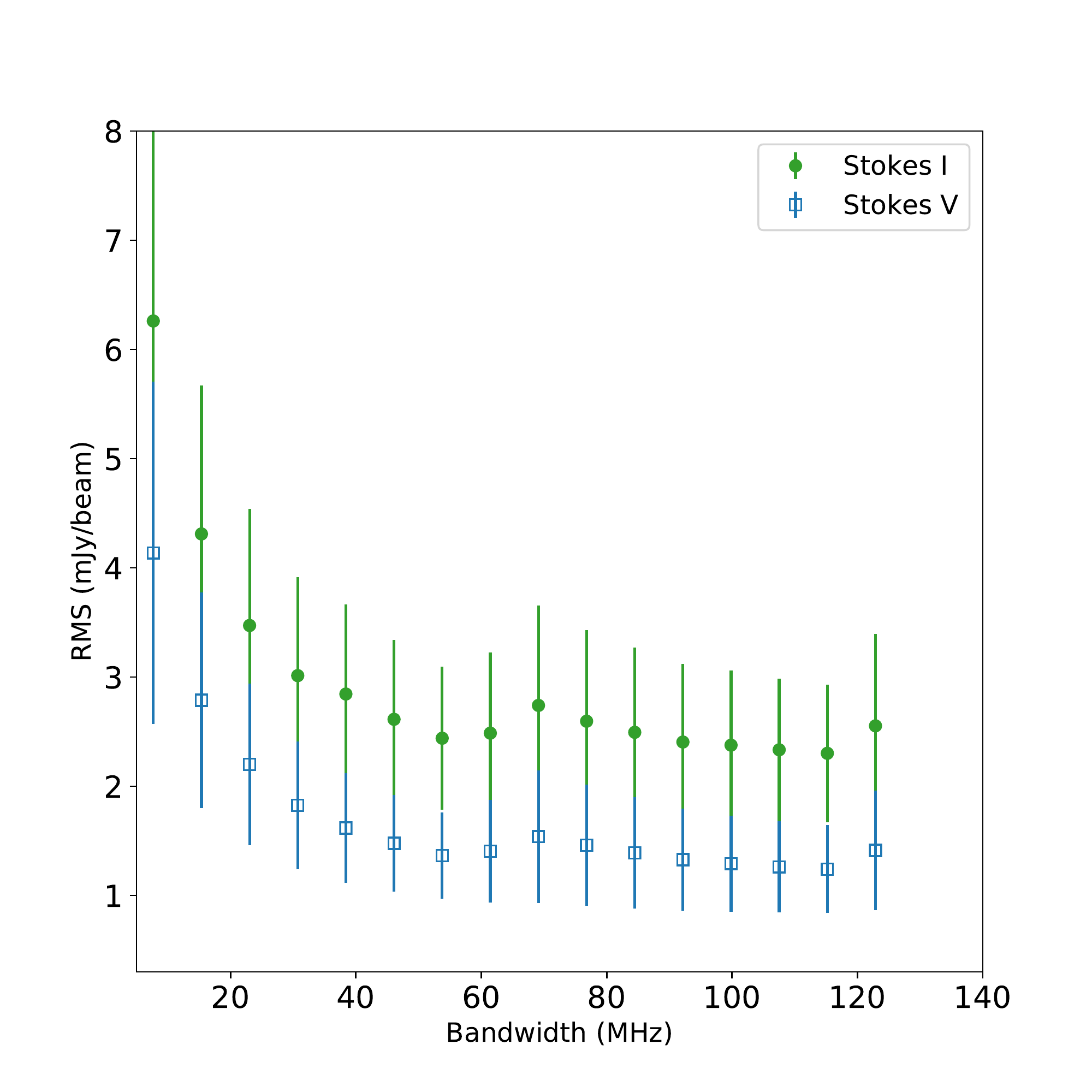}
\caption{The average rms (within a central 18 x 18~degrees box) for combined spectral images for the LoBES field 1 as a function of the combined bandwidth. The green circles represent the Stokes I values and the blue open squares the Stokes V values; the uncertainty for each point is the standard deviation of the rms within the region. }\label{fig:rmsbw}
\end{center}
\end{figure}

\subsection{Wide-band Image creation}

To determine the set of spectral images to use to create the wide-band image, starting with the highest frequency spectral image, we iteratively combined subsets of the spectral images (proceeding to lower and lower frequencies). To join the spectral images, we first convolve the subset of images to the resolution of the lowest frequency image, and then mosaic the images using \textsc{swarp}, as described in Section \ref{sec:mosaic}. Using the root-mean-squared (rms) as an estimate of the image noise, we then measured the average rms within a 18 x 18~degrees central region of the produced wide-band image. The evolution of the rms as a function of combined bandwidth for the LoBES 1 field is shown in Figure \ref{fig:rmsbw} for both Stokes I and V images (the other fields have a similar rms evolution). Note that the rms estimates for the Stokes V images are shown as an estimate of the expected thermal noise for the images. There is a slight offset between the Stokes I and V values, however for most of the integrated images the values agree within the 1$\sigma$ error bars. The observed offset is the result of significant variation in the rms around bright sources in Stokes I; the Stokes V image does not include any bright sources and the rms is much more uniform across the image.

\begin{figure*}[t!]
\begin{center}
\includegraphics[scale=0.85]{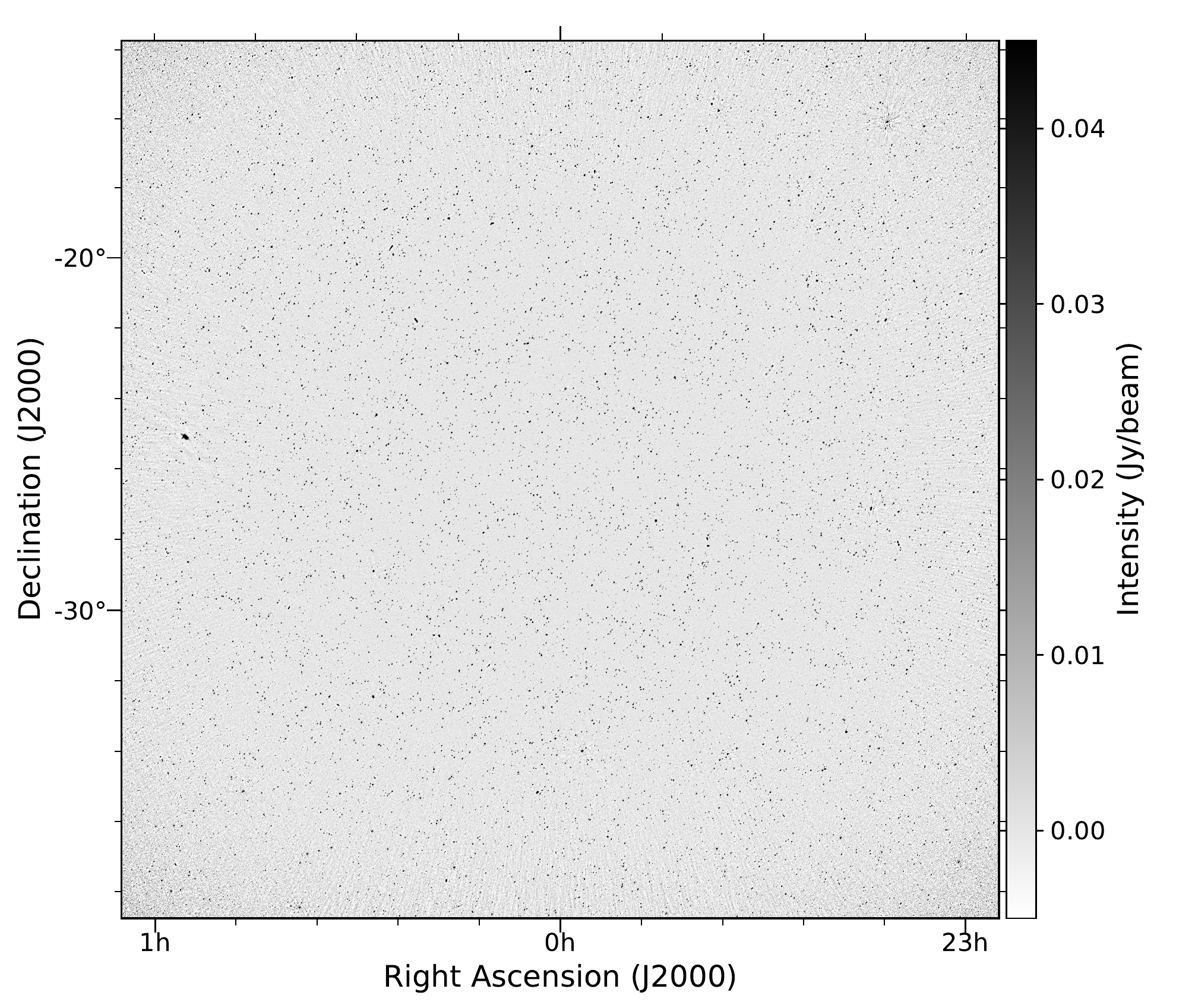}
\caption{Central 25 x 25~degree region of the LoBES field 1 wide-band image highlighting the high image quality of MWA phase 2 extended. The noise at the edge of the field, due to the primary beam attenuation, is apparent.}\label{fig:eor0-deep}
\end{center}
\end{figure*}

The Stokes I image rms is first minimised in the five LoBES fields around 60~MHz of combined bandwidth; this corresponds to the spectral image range of 170~--~230~MHz. Continuing to combine spectral images beyond 60~MHz of bandwidth is not beneficial as there is no improvement in the image rms, and the image resolution will continue to degrade as we convolve to image resolutions of the lower frequency images added to the subset. Figure \ref{fig:eor0-deep} shows the inner 25 x 25~degrees of the LoBES field 1 wide-band image. The better point spread function (PSF) characteristics of the MWA Phase II Extended configuration are evident, as many of the image artefacts observed in Phase I data around bright sources in this field are significantly reduced (see Figure 1 from \citet{Offringa:2016} for comparison).

\begin{figure*}[t!]
\begin{center}
\includegraphics[scale=0.48]{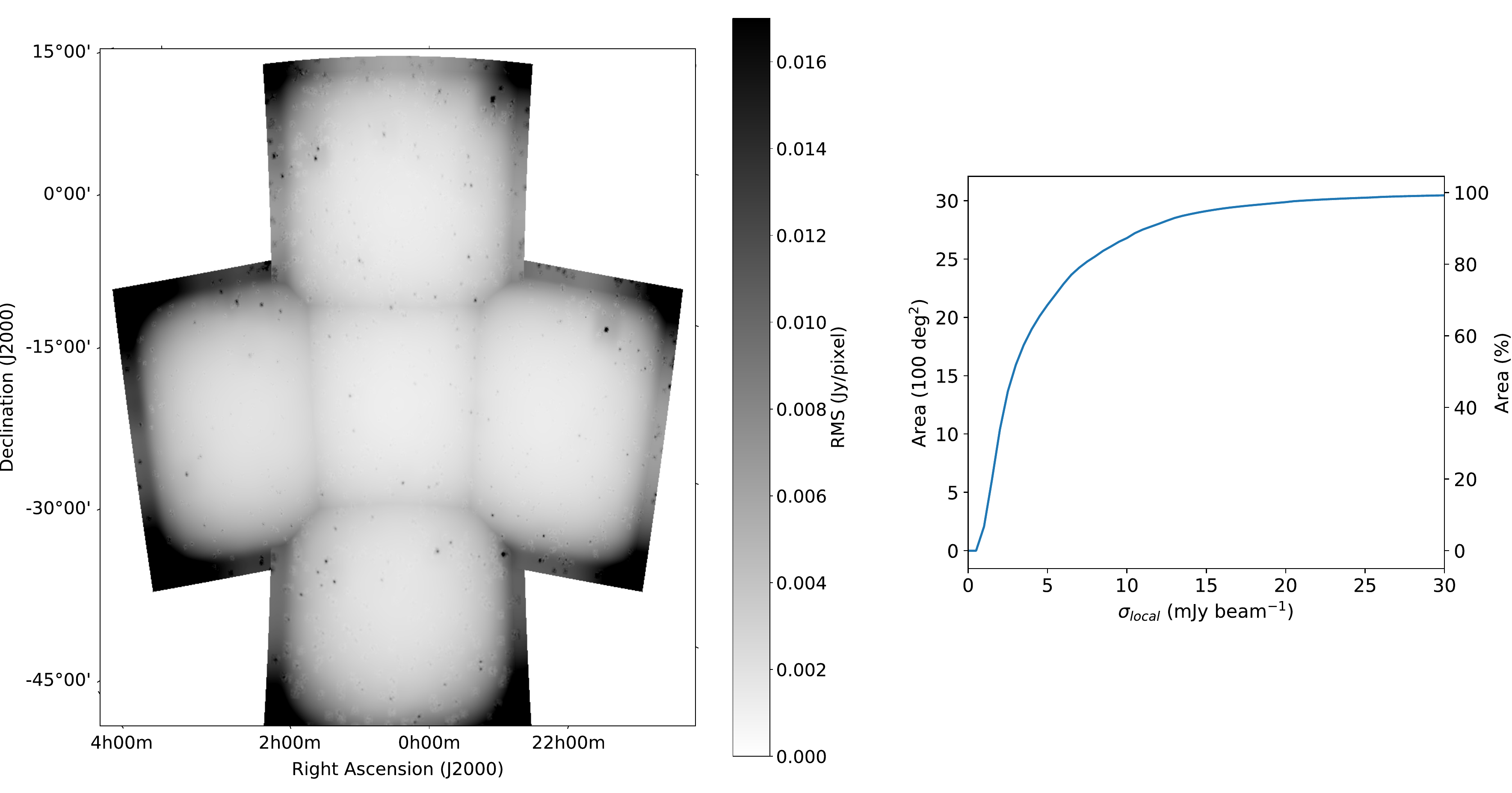}
\caption{The left-hand figure shows the sensitivity coverage over the full survey area presented here. Note that the rms is not uniform across the surveyed sky area, and increases toward the edges of the survey fields and around bright sources in the fields. The right-hand figure shows the cumulative area coverage (and percentage) that has an rms less than a given value. 
}\label{fig:rmsvar}
\end{center}
\end{figure*}

We created an overall rms map of the wide-band images for the LoBES fields (left panel of Figure \ref{fig:rmsvar}) by mosaicking the noise maps of each field generated via the source finder \textsc{pybdsf} (Python Blob Detector and Source Finder; \citet{Mohan:2015}). To be consistent with the source selection for overlapping fields (as outlined in Section \ref{sec:srcselect}), when mosaicking the noise maps for each field we choose the \textsc{swarp} combine type `MIN', which selects the minimum pixel value as the output pixel. For the rms mosaic, we use a mollweide projection, which is an equal area projection. The rms variation across the LoBES fields is apparent in the left panel of Figure \ref{fig:rmsvar}; this variation is caused by the attenuation of the primary beam at the field edges and residual noise around bright sources within the fields. The average rms in a box region of size 14 x 48~degrees, centred at RA~(J2000)~0\textsuperscript{h}, Dec~(J2000)~$-$27$^{\circ}$, in the north-south direction has an average rms of 2.1~mJy~beam$^{-1}$. A similar box in the east-west direction has the same average rms. The right panel of Figure \ref{fig:rmsvar} shows the area, and corresponding percentage, of the image that has an rms value less than a given value. Roughly 85$\%$ of the area of the survey has an rms $<$~9.0~mJy~beam$^{-1}$.

\subsection{Reference Source Catalogue}\label{sec:refCat}

We used \textsc{pybdsf} to find and fit sources in the wide-band image for the reference catalogue. \citet{Procopio:2017} showed that errors associated with mismodelling bright, extended sources contributes the most to residual foreground power. Thus it is important to use a source finder that can robustly fit these more complex sources. The quality of source fitting performed by various source finders has been compared in previous tests of radio survey data \citep[e.g.][]{Hopkins:2015}. In these tests, \textsc{pybdsf} is shown to perform well, as compared to other source finders, when fitting extended and complex sources \citep{Hopkins:2015, Hale:2019}.

Using \textsc{pybdsf}, the background noise across each of the wide-band images was estimated using sliding box sizes of 735 pixels. To more accurately capture the increased local rms in regions surrounding high signal-to-noise sources ($\geq$150$\sigma$), we decrease the size of the sliding box to 38 pixels near such sources. To find and fit sources, \textsc{pybdsf} first identifies all pixels in the image greater than a set pixel threshold. Starting from each identified pixel, all contiguous pixels higher than a set island threshold are established as belonging to one island. The islands are then fitted with multiple Gaussians and nearby Gaussians within an island are grouped into sources. For the LoBES wide-band images, we used the default 5$\sigma$ pixel threshold for detection and 3$\sigma$ island threshold. Additionally, we used the \textsc{pybdsf} parameter $group\_tol$, which controls how Gaussians within the same island are grouped into sources. We set this parameter to 3 to allow for larger sources to be formed. This value balances grouping all Gaussians in a single island into one source and merging too many separate sources, and identifying the island Gaussians as separate sources, separating components of large, extended sources. This value was optimised through trial and error testing using the interactive mode of \textsc{pybdsf}. 


Since the pixels in the wide-band images contain contributions from a set of images covering close to a full hour of observation, any residual offsets due to ionospheric activity during these observations will cause the PSF of the wide-band image to blur and vary across the FOV. To address this issue we use \textsc{pybdsf} to estimate the spatial variance in the PSF and correct for its effects. Using a set of unresolved, bright ($>$~10$\sigma$) sources, \textsc{pybdsf} tessellates the image using a Voroni tessellation to produce a set of tiles. The bright, unresolved sources within a tile are then used to calculate the PSF for that tile and each tile is assumed to have a constant PSF. The spatial variation of the PSF is then quantified based on the tile PSF values and interpolated across the whole image. The deconvolved source sizes are then adjusted by \textsc{pybdsf} to include the PSF variation as a function of position. 

For each of the five wide-band images, \textsc{pybdsf} produces a reference source Catalogue with parameters relating to the overall source flux densities, sizes, and positions, a reference Gaussian catalogue with information about the Gaussian components (flux densities, positions, sizes, orientations) used to fit each source. The uncertainties on the fitted parameters are computed following \citet{Condon:1997}. Additionally, each of the sources identified and fit by \textsc{pybdsf} is given a source code (the `S\_Code' column in outputted catalogues). Sources fit by multiple Gaussians are given a `M' source code, sources fit by a single Gaussian, a `S' source code, and single-Gaussian sources that lie within the same island as another source are given a `C' source code. From \textsc{pybdsf} we also have rms maps, residual maps, and PSF maps associated with each field. The total number of sources found in the five LoBES wide-band images is 100851. Note that this initial number contains edge sources, sources common to more than one LoBES field, spurious emission or artefacts, and large extended sources whose individual components appear as separate sources in the \textsc{pybdsf} source catalogues; these issues will be handled in the following section.

\subsubsection{Final source selection}\label{sec:srcselect}
Each of the five wide-band images have hard edges where a small number of sources should be omitted. These sources may still be detected by \textsc{pybdsf} at the edge of the image, but these sources are likely to be incomplete or have erroneous flux densities and shapes. Thus we have removed sources that have fitted sky positions, that when converted to pixel values are within 10 pixels of the edge of the image. Generally less than 1$\%$ of the total sources from each reference catalogue are removed during this step. 

Adjacent LoBES fields overlap so that some sources can be present in multiple wide-band images. To avoid double counting sources we need to identify and remove copies of the same source in multiple source catalogues. To do this we first use the Starlink Tables Infrastructure Library Tool Set (\textsc{stilts}; \citet{Taylor:2006}) to cross-match each wide-band source catalogue with the catalogues of their adjacent fields (for example LoBES field 2 with fields 1, 3, and 5). We identify all matched sources within 80~arcseconds between two LoBES fields using the \textsc{tskymatch2} command.  For the double entries, we discarded one of the two by directly comparing the Island rms values reported by \textsc{pybdsf} during the initial source finding, choosing to keep the source with the smallest rms. Removing double-counted sources removes 20$\%$ of the total sources from the reference catalogues. 

We visually inspected a sub-set of the catalogue sources to identify spurious imaging artefacts that were identified as sources, and to find sources that are actually individual components, such as radio lobes, of large extended sources. The source catalogue was first separated into three catalogues each containing one of the three types of sources (S, M, or C). For each of the M and C sources we created 0.5~x~0.5~degree sub-images centred on the source from the appropriate LoBES wide-band image. For the M sources we overlaid the sub-images with markers to indicate the position, source size and orientation of the source. The C sources similarly were marked with the location and size of the source of interest but we also included additional markers to identify all other sources located in the source island. Doing this for all five of the LoBES fields created 8500 sub-images to inspect by eye. Visually inspecting the M and C source subimages revealed 91 individual sources that are actually components of a large extended source or imaging artefacts. We remove these sources and their Gaussians fits from the source and Gaussian catalogues and the identified large extended sources are refit using \textsc{pybdsf} parameters better suited to extended emission (see Section \ref{sec:refit}). 

Given that the majority of the sources in the LoBES field reference catalogues are S sources, and that for each field there are tens of thousands of these sources, it is unreasonable to visually inspect each of these sources. To get a sense of where bright imaging artefacts might lie in the images, we inverted the LoBES field 1 wide-band image (by taking the negative of every pixel) and running \textsc{pybdsf} on the inverted image using the same fitting parameters used to fit the original wide-band images. We expect imaging artefacts around bright sources will have negative counterparts that are detectable in the inverted image. We find that the majority of the inverted sources lie within 5 arcminutes of sources with integrated flux densities greater than 5~Jy. To account for any noise variation between the LoBES fields we choose only S sources in the wide-band (non-inverted) source catalogue that lie within 5 arcminutes of a source that has an integrated flux density greater than 2~Jy. This will select S sources that are most likely to be identified as spurious imaging artefacts. This selects 791 S sources; we create sub-images centred on each of the identified sources and by visually inspecting these sources, we identified 38 sources to remove from the source and Gaussian catalogues.

Overall a total of 80824 sources (comprising of 88381 Gaussian components) are identified in this first half of the LoBES survey -- excluding the forty-five large extended sources we choose to re-fit in Section \ref{sec:refit}. We collect these remaining identified sources from each LoBES field reference catalogues into a single final source and Gaussian catalogue. For each source, the final source catalogue contains information about the peak intensity, integrated flux density, position, orientation, convolved and deconvolved shapes, and the associated errors for each of the parameters. Similarly, the final Gaussian catalogue contains this information but for each Gaussian fit to the identified sources. A column was added to these catalogues to include the LoBES field from which each source was found. We also add identifying columns to the final source and Gaussian catalogues, where each source is given a unique name and source ID number. These are used in the final Gaussian catalogue to associate Gaussian components to a single identified source. We will refer to these two catalogues as the General Wide-band Source and General Wide-band Gaussian catalogues.

\subsubsection{Re-fitting large extended sources}\label{sec:refit} 

The forty-five large extended sources identified in the preceding Section are re-fitted using \textsc{pybdsf} parameters that are more suited to fitting dimmer extended emission\footnote[3]{For more details see \url{https://www.astron.nl/citt/pybdsf/examples.html##image-with-extended-emission}}. We create 0.3~x~0.3~degree sub-images centred on each of these sources from the appropriate LoBES wide-band image. We then fit each of these sub-images using the interactive mode of \textsc{pybdsf}, varying the pixel and island thresholds to obtain an island that enclosed all significant emission, and $group\_tol$ to group all fitted Gaussians into a single source. We also set the $rms\_map$ parameter to `false' and the $mean\_map$ parameter to `const', forcing \textsc{pybdsf} to use a constant mean and rms values across the sub-image; these settings account for the background rms map likely being biased high in regions where extended emission is present. We set $flag\_maxsize\_bm$ to 50 to allow for large Gaussians to be fit, and $atrous\_do$ to `True' to fit Gaussians of various scales to the residual image to recover extended emission missed in the standard fitting. The source catalogue for each of the forty-five sources was combined into a single catalogue (similarly for the individual Gaussian catalogues) and we will refer to these two catalogues as the LG-Extended Wide-band Source and LG-Extended Wide-band Gaussian catalogues; these two catalogues contain the same information as the General Wide-band Source and Gaussian catalogues.

\subsection{Spectral Catalogue}\label{sec:specmeas} 

To measure the flux densities for all of the sources in the reference catalogue, we use the priorised fitting algorithm within \textsc{aegean} \citep{Hancock:2018}. This algorithm takes an input catalogue of sources that contains the source positions and morphologies and measures the flux density of each source within a supplied image. We use the final reference LG-Extended and the General Wide-band Gaussian catalogues, detailed in Section \ref{sec:refCat}, as the input catalogues for the priorised fitting and the supplied images are the sixteen spectral images associated with each LoBES field. Using the `LoBES$\_$FIELD' columns in each of the reference catalogues, we only fit sources in the spectral images associated with the LoBES field they were identified in (i.e. LoBES field 1 sources are only priorised fit in LoBES field 1 spectral images). We re-format the Gaussian catalogues to have a similar formatting to that of an \textsc{aegean} catalogue and include an additional column with a unique identifier for each of the source Gaussians (the `UUID' column in an \textsc{aegean} catalogue). To account for PSF differences between the input catalogue and the image, we use the PSF images created by \textsc{pybdsf} for each of the wide-band images, to supply \textsc{aegean} with the local PSF information for each of the reference catalogue sources. 

Before fitting, \textsc{aegean} deconvolves sources by the local catalogue PSF and convolves them with the local PSF of the supplied image. Sources are then re-grouped to create islands of sources based on their positions and morphologies. Sources are grouped together if they overlap at the half-power point of their respective Gaussian fits. \textsc{aegean} then fits each of the identified islands, jointly fitting sources grouped together. By jointly fitting grouped sources, \textsc{aegean}'s priorised fitting accounts for biases in the fitted flux densities due to source blending \citep{Hancock:2018}. Different fit parameters can be allowed to vary during the priorised fitting -- during our fitting of the LoBES spectral images we choose to only fit for the flux density of each fit Gaussian, while fixing its shape and position. The priorised fitting creates an output catalogue with the fixed and fitted parameter information for each of the source Gaussians, as well as the supplied unique identifier (`UUID') from the input catalogue. We use the `UUID' to match and collect the spectral information for each of the fit Gaussians into a single spectral Gaussian catalogue. To get the total integrated flux density for each source within each frequency band, we sum the integrated flux densities of the fit Gaussians for each source at each frequency and use the position and source size from the General and LG-Extended Source catalogue to create spectral source catalogues. 

\subsection{Fitting spectral models}\label{sec:specfit} 
Using the spectral source catalogues, we performed spectral modelling for all LoBES sources that are measured to have at least six spectral measurements that are greater than 4$\sigma$; this includes 78~$\%$ of the total LoBES sources. This selection is to ensure that the sources have a sufficient number of significant spectral measurements to do the modelling. To account for the overall flux density scale uncertainty for the LoBES measurements, the uncertainty for each measurement is calculated to be the quadrature sum of the \textsc{aegean} fitting uncertainty and a 5$\%$ flux density scale error as calculated in Section \ref{sec:err}. 

To include spectral information from archival multifrequency radio catalogues in the spectral modelling, we used \textsc{PUMA} to cross-match the spectral General Source catalogue with VLSSr, SUMSS, and NVSS. However, for the LG-Extended sources we were concerned that each source could be composed of multiple `source' entries in these archival catalogues. To ensure correct source association for the LG-Extended sources, we downloaded VLSSr, NVSS, and SUMSS images for each of these sources from the SkyView Virtual Observatory \citep{McGlynn:1998}, performed an \textsc{aegean} priorised fit in each image using the LG-Extended Gaussian catalogue, and summed the measured flux densities of the fit Gaussians to get the total integrated flux density for each source in their respective images. 

\subsubsection{Spectral models}
To capture the spectral shape for each LoBES source, we fit two different models. Radio sources often exhibit simiple spectra that can be approxmiated by the standard non-thermal power-law model, where the flux density, $S_{\nu}$, at frequency $\nu$ is given by
\begin{equation}\label{eq:pwlaw}
S_{\nu} = a \left(\nu/\nu_{o}\right)^{\alpha},
\end{equation}
here $a$ is the amplitude of the synchrotron spectrum in Jy and $\nu_{o}$ is the reference frequency where we used 160~MHz in this analysis. Yet at low radio frequencies, source spectra are found to be more complex, with sources showing more curved spectra \citep[e.g.][]{Laing:1980, Blundell:1999, Duffy:2012, Marvil:2015, Callingham:2017, Galvin:2018} until a turnover frequency, below which they become inverted. Synchrotron self-absorption or free-free absorption is typically thought to be responsible for the observed spectral turnover \citep{Callingham:2015, Callingham:2017}. To capture potential curvature in the spectra of the LoBES sources, we also fit each source using a curved power-law model of the form
\begin{equation}\label{eq:cpwlaw}
S_{\nu} = a \left(\nu/\nu_{o}\right)^{\alpha} e^{q \ln (\nu/\nu_{o})^2}.
\end{equation}
Here $q$ is the spectral curvature, where values of $\mid$~$q$~$\mid$~$>$~0.2 represent significant curvature, and the spectral curvature flattens toward a standard power-law as $q$ goes to zero. 

\subsubsection{Fitting and selection}
Following the method outlined by \citet{Callingham:2015, Callingham:2017} and \citet{Galvin:2018} we use a Bayesian model inference routine to assess the parameter values of the two spectral models fit to each of the catalogue sources. This routine samples the posterior probability distribution functions of the model parameters using an affine invariant Markov Chain Monte Carlo algorithm \citep{Goodman:2010}, implemented via the \textsc{emcee} python package \citep{Foreman:2013}. Final model parameters are chosen such that they maximise the log likelihood function under physically sensible uniform priors. Here the log likelihood function is given by
\begin{equation}\label{eq:loglike}
\ln \mathcal{L}(\theta) = -\frac{1}{2} \sum_{n} \left[\frac{(D_n - f(\theta))^2}{\sigma_{n}^2} + \ln(2\pi\sigma_{n}^2)\right],
\end{equation}
where $D$ and $\sigma$ are vectors containing a set of $n$ flux density measurements and their associated uncertainties, and $f(\theta)$ is the model optimised using the parameter vector $\theta$. 

Equation \ref{eq:loglike} assumes that the measurements are independent with normally distributed errors. However, the 7.68~MHz sub-band measurements have correlated errors, violating this underlying assumption. Similar to GLEAM, this correlation is introduced through a combination of the primary beam correction, the absolute flux density scaling and ionospheric corrections, self-calibration, and multi-frequency synthesis performed on the full 30.72~MHz observing bandwidth before splitting it into the four narrower sub-bands. As some of these effects have a direction-dependent component, the degree of correlation between the sub-bands varies as a function of position. In order use the flux density measurements from the LoBE survey in combination with those from other radio surveys for the spectral modelling, the correlation between the LoBE survey sub-bands needs to be accounted for with an appropriate covariance matrix. 

Calculating the exact form of the covariance matrix describing the correlation between the LoBES points is not possible. Following \citet{Callingham:2015, Callingham:2017}, we can approximate the correlation using a Mat\'ern covariance function \citep{Rasmussen:2006}. This type of covariance function produces a stronger correlation between flux density measurements closer together in frequency than further apart. The parameterised Mat\'ern covariance function, $k$, we use in the spectral modelling is given by
\begin{equation}\label{eq:matern}
k(r) = N^2 \left(1 + \frac{\sqrt{3}r}{\gamma}\right)e^{\left(\frac{-\sqrt{3}r}{\gamma}\right)},
\end{equation}
where $r$ is the difference in frequency between pairs of flux density measurements, and $N$ and $\gamma$ are quantities constrained by \textsc{emcee}. We use the python package \textsc{george}\footnote[3]{https://github.com/dfm/george} \citep{Ambikasaran:2015} to implement the Mat\'ern covariance function and supply the log likelihood function of a model given the parameter vector $\theta$ for only the LoBES measurements. The log likelihood function from \textsc{george} was summed with the log likelihood function from equation \ref{eq:loglike} for the independent flux density measurements and parameter vector $\theta$. 

In this modelling, we use uniform priors enforcing a range of values that the model parameters are allowed to take. Throughout our model fitting we ensure that the spectral index remains in the range of -3 to 3, covering the broad range of measured values in the literature. Additionally, we force the amplitude values $a$ to be positive under the assumption that the flux densities are the result of a positive emission process. For the Mat\'ern covariance parameters $N$ and $\gamma$ we make no assumption about their values and set the priors broadly enough to cover all the LoBES data.

In the Bayesian framework we can select between two equally likely models, $M_1$ and $M_2$, by comparing the Bayesian evidence values for each model evaluated for a common dataset. The evidence value, $Z$, is given by 
\begin{equation}\label{eq:ev2}
Z = \iint\cdots\int \mathcal{L}(\theta) \Pi(\theta) d(\theta)
\end{equation}
where $\Pi(\theta)$ is the prior probability distribution and the dimensionality of the integration depends on the number of model parameters. The evidence value represents the average likelihood over the prior volume for a given model, and favours models with high likelihood throughout the prior parameter space. Therefore a simpler model containing few parameters will have a larger evidence value than a more complex model with a larger parameter space, except if the complex model is a significantly better fit to the data. We calculate the evidence values for the two different models using the \textsc{dynesty} python package \citep{Speagle:2020}, which uses a nested sampling method \citep{Skilling:2004, Skilling:2006} to obtain an estimate of the $Z$ values. The prior volume searched by \textsc{dynesty} is informed by the results of the fitting using \textsc{emcee}.

\begin{table}
\caption{The LoBE survey properties}
\begin{tabular}{lc}
\hline
Total sky coverage & 3069~degree$^2$ \\
Number of sources & 80824 \\
Completeness at 10.5~mJy     & 70$\%$\\
Completeness at 50~mJy     & 94$\%$ \\
Completeness at 100~mJy    & 98$\%$\\
Completeness at 1 Jy       & 100$\%$ \\
rms (mean)              & 2.1~mJy~beam$^{-1}$ \\
Wide-band image resolution & 77.5~arcsec \\
R.A. offset &$-$0.04$\pm$1.48~arcsec\\
Dec. offset &$-$0.03$\pm$1.53~arcsec \\
flux density scale uncertainty & 5$\%$ \\
\hline
\end{tabular}\label{tab:stats}
\end{table}

Given the evidence values, $Z_1$ and $Z_2$, for models $M_1$ and $M_2$, we can use the ratios of the evidences to select the model that better fits our flux density measurements. In this paper we perform this comparisons in log space such that,
\begin{equation}\label{eq:ev1}
\Delta \ln (Z) = \ln(Z_2) - \ln(Z_1)
\end{equation}
If $\Delta \ln (Z)$~$\geq$~3, model $M_2$ is strongly favoured over $M_1$; model $M_2$ is only moderately favoured over $M_1$ if 1~$<$~$\Delta \ln (Z)$~$<$~3; and if $\Delta \ln (Z)$~$<$ 1, preference of one model over the other is inconclusive \citep{Kass:1995}. We chose $M_2$ to be the curved power-law model, and find that for 20$\%$ of the modelled sources the curved power-law is strongly favoured. 


\section{Final Catalogue Properties}

The final LoBES catalogue combines the positional and shape information from the General and LG-Extended Wide-band catalogues with the spectral information outlined in Sections \ref{sec:specmeas} and \ref{sec:specfit}. The sources that do not have sufficient spectral information to carryout modelling are included in the final catalogue but we only include their measured values and associated uncertainties. The catalogue associated with the results of this first paper will be available upon request. A final catalogue containing the full LoBE survey coverage (including the sources presented here) will be released with paper 2. 

\begin{figure}[t!]
\begin{center}
\includegraphics[width=\columnwidth]{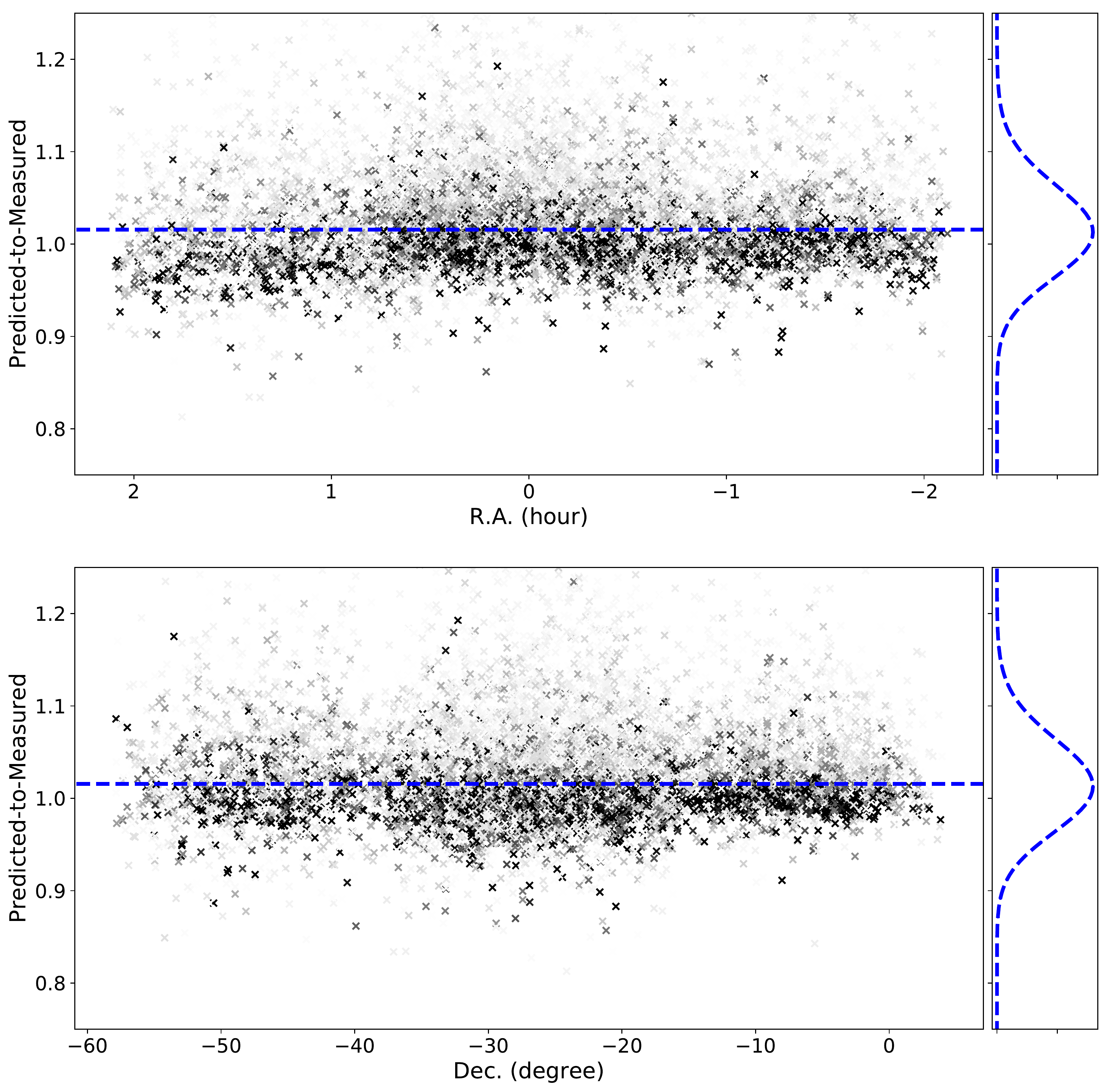}
\caption{Ratios of the overall flux density scale in the LoBES catalogue at 189~MHz as compared to the predicted flux densities via cross-matching to other catalogues (the plots are similar for the other spectral images). We show these ratios as a function of both Right Ascension (top) and Declination (bottom). The grey-scale represents the signal-to-noise ratio of the selected sources, with darker colors associated with the highest values. On the right of each figure we also show the weighted log-Gaussian fit to the spread in the ratio values. We take the standard deviation of the fit log-Gaussian to be the systematic uncertainty in the flux density scale. For all spectral images the uncertainty is found to be 5$\%$.}\label{fig:flxunc}
\end{center}
\end{figure}

\subsection{Error estimation}\label{sec:err}
In the following we summarise how we estimated the uncertainties in the LoBES external flux density scale and position measurements. To calculate the systematic uncertainty in the LoBES spectral measurements, we first identify unresolved, isolated sources with a signal-to-noise ratio $\geq$8 in each of the 16 spectral catalogues. We cross-match these sources with NVSS, SUMSS, GLEAM, and VLSSr using \textsc{PUMA}. Again we prioritise the NVSS positions over the other catalogues. Carrying out a similar analysis as that used to measure and correct for the flux density scale variation in the individual LoBES images, we estimate the ratio of the measured to predicted flux density for each of the bright sources found in the spectral images. We fit a weighted log-Gaussian to the distribution of ratios, where the weight for each source is taken to be the signal-to-noise ratio, and take the standard deviation of this fitted Gaussian as the uncertainty in the flux density scale at that frequency.  For all spectral images, the systematic uncertainty is found to be 5$\%$. We show the distribution of flux density ratios for all five of the 189~MHz images in Figure \ref{fig:flxunc} -- the other spectral images look similar. Here the grey-scale represents the signal-to-noise ratio for each of the sources, with the darkest colors associated with the highest signal-to-noise ratios. We additionally show the fitted log-Gaussian in the right-hand panel of this figure. 

\begin{figure}
\begin{center}
\includegraphics[width=\columnwidth]{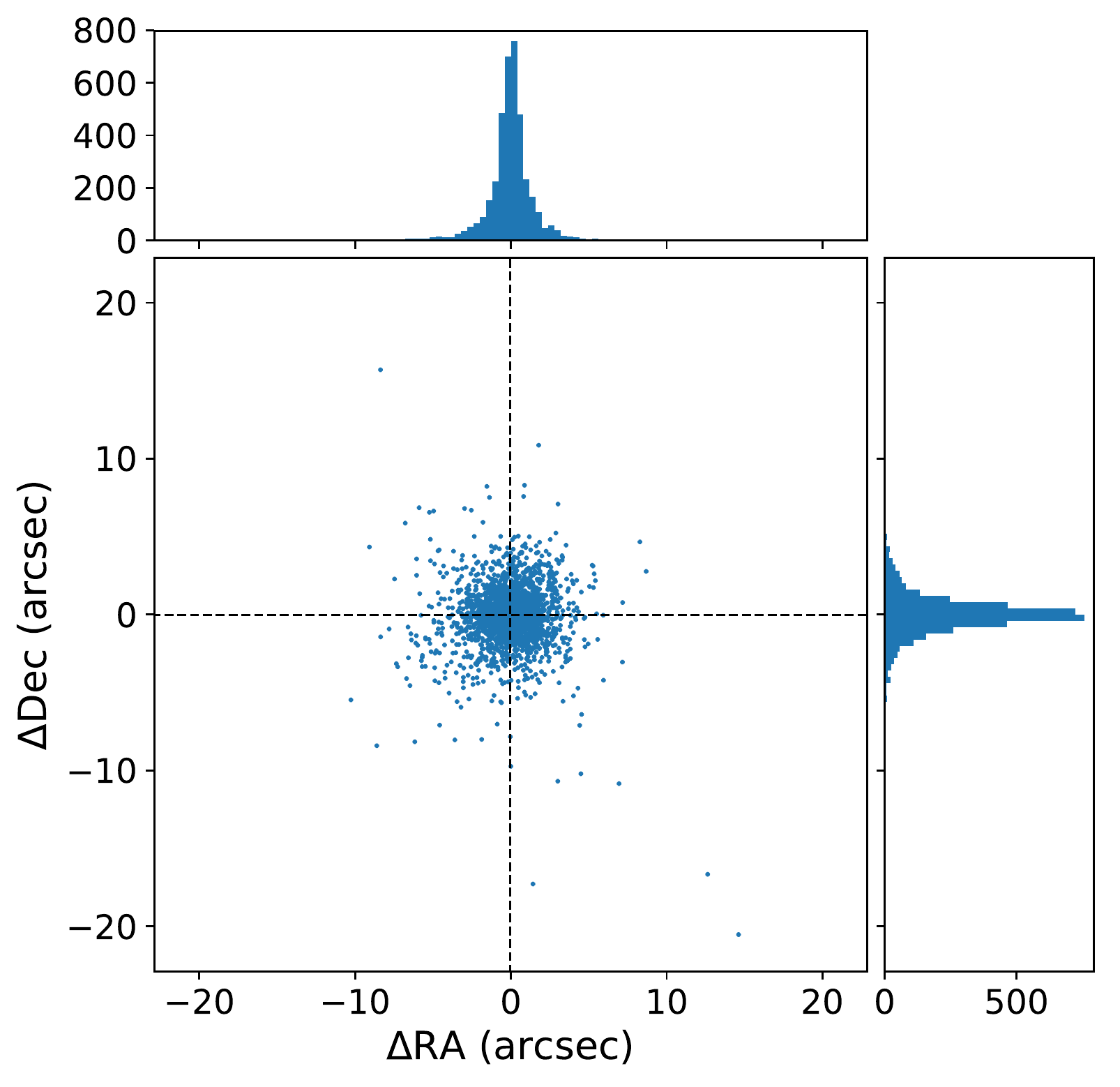}
\caption{Positional offsets in RA and Dec as calculated from the updated source position as reported from a cross-match with NVSS, SUMSS, GLEAM and VLSSr using \textsc{PUMA}. We also show histograms of the differences in RA (top) and Dec (right), which are tightly centred on zero. The dashed line represents the average offset in both cases, which is smaller than 99$\%$ of the source fitted positional uncertainties. }\label{fig:astrometry}
\end{center}
\end{figure}

Similarly, to measure the astrometric uncertainty in the LoBES catalogue, we identify unresolved, isolated sources with a signal-to-noise ratio $\geq$8 in the General Wide-band catalogues. We cross-match these sources with NVSS, SUMSS, GLEAM and VLSSr using \textsc{PUMA} and calculate the difference in the sky position between the original LoBES position and the updated position reported by \textsc{PUMA}. The distribution of these positional differences are shown in Figure \ref{fig:astrometry}. For these sources the average offset in Declination and Right Ascension is consistent with zero. These offsets are smaller than the \textsc{pybdsf} fitted Gaussian uncertainties for 99$\%$ of the sources in the catalogue. Given the small scatter in these offsets, and their relative size to the Gaussian fit uncertainties, we do not make any corrections for the offsets. 

\subsection{Spectral indices}
\begin{figure}[t!]
\begin{center}
\includegraphics[width=\columnwidth]{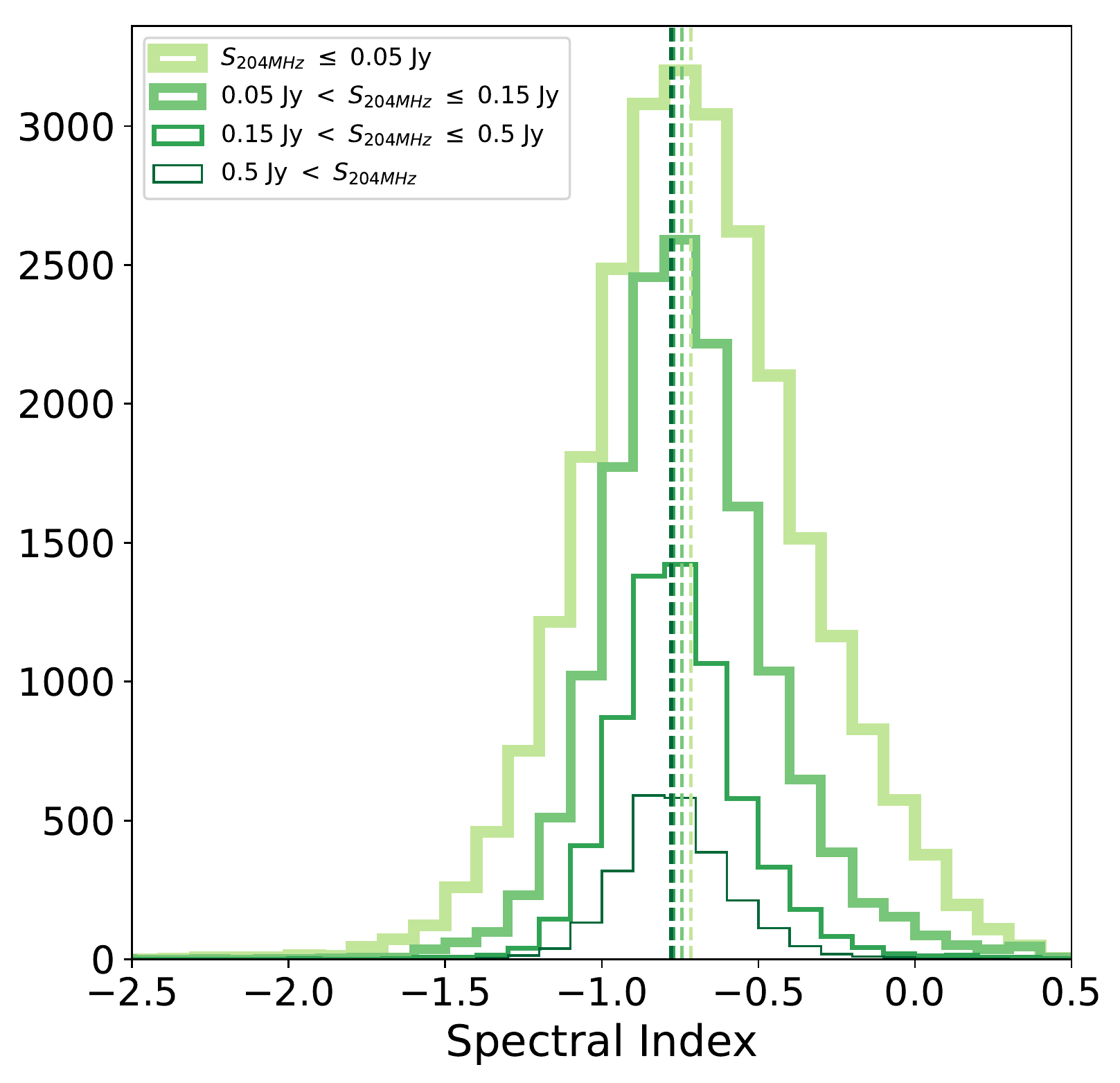}
\caption{The spectral index distribution for sources best fit by a power-law in Section \ref{sec:specfit}. The sources are grouped by their measured flux density at 204~MHz. From lightest shade of green to darkest the bins are: flux densities $\leq$ 0.05~Jy;  flux densities between 0.05 -- 0.15~Jy; fluxes between 0.15 -- 0.5~Jy; and flux densities greater than 0.5~Jy. The corresponding dashed lines indicate the associated median spectral indices for each flux bin. }\label{fig:sindx_hist}
\end{center}
\end{figure}

To assess the quality of the spectral modelling in Section \ref{sec:specfit} we can compare the average spectral index for the sources fit by a power-law in the LoBE survey, to spectral indices measured by other radio surveys. We identify 50673 sources whose spectral properties are best fit by a standard power-law. The histogram of spectral indices for these sources in Figure \ref{fig:sindx_hist}. Here we have broken the sources into four flux density bins: $S_{204~\mathrm{MHz}}$ $\leq$ 0.05~Jy (26187 sources); 0.05~Jy $<$ $S_{204~\mathrm{MHz}}$ $\leq$ 0.15~Jy (15327 sources); 0.15~Jy $<$ $S_{204~\mathrm{MHz}}$ $\leq$ 0.5~Jy (6670 sources); $S_{204~\mathrm{MHz}}$ $>$ 0.5~Jy (2489 sources). The associated median spectral indices for these flux density bins are $-$0.72~$\pm$ 0.22, $-$0.75~$\pm$ 0.16, $-$0.77~$\pm$ 0.12, and $-$0.78~$\pm$ 0.11, respectively. 

Several other papers have reported spectral indices for sources identified at $\sim$100~MHz. Our median spectral indices are in agreement with these previous results. GLEAM measured the spectral indices within the frequency coverage of that survey, breaking the sources into flux density bins. The bottom two flux density bins used by GLEAM are similar to those we use here; they report spectral indices of $-$0.78~$\pm$ 0.2 and $-$0.79~$\pm$ 0.15 in these bins. While our top flux density bin includes both the top flux density bins used in GLEAM, the spectral indices for this range of flux densities ($>$ 0.5~Jy) are in agreement, with GLEAM reporting a spectral index of $-$0.83~$\pm$ 0.12 and $-$0.83 $\pm$ 0.11 in these two bins. Similarly, our results agree with \citet{Williams:2016} who reported an average spectral index of $-$0.79~$\pm$ 0.01 between 150 and 1400~MHz for sources located in the Bo\"otes field; \citet{deGasperin:2018} used the TGSS ADR1 and NVSS catalogues to create a large area spectral index map, they report median spectral indices that span the range of $-$0.611 in their lowest flux density bin (0.0025 -- 0.05~Jy) to $-$0.81 in the highest flux density bin (0.5 -- 500~Jy) ; finally \citet{Heald:2015} used the LOFAR High Band Antenna to measure a median index of $-$0.77 (between 120 -- 160~MHz). 

\begin{figure}[b!]
\begin{center}
\includegraphics[width=\columnwidth]{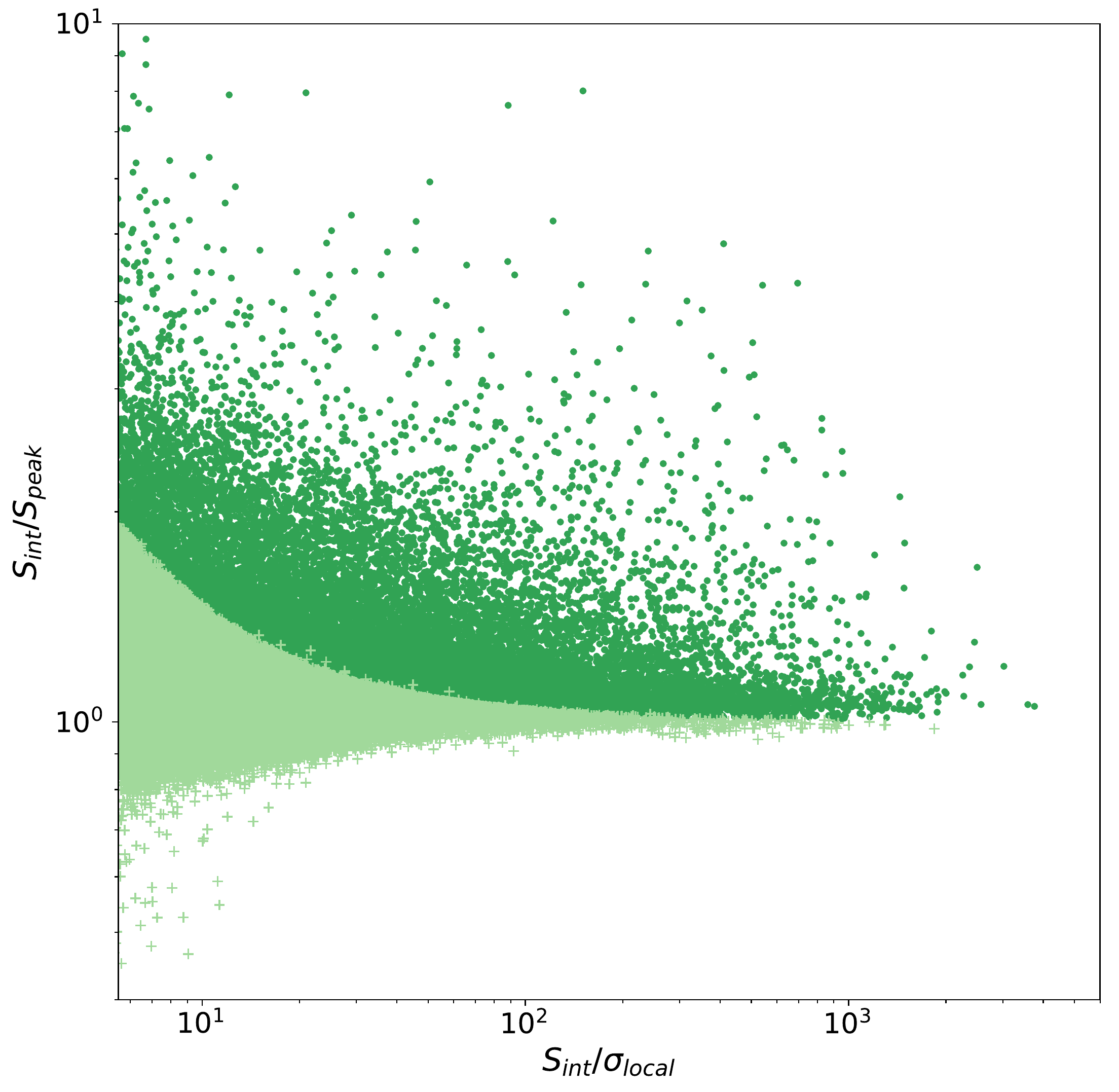}
\caption{Ratios of the integrated to peak intensity as a function of the signal-to-noise ratio of the source. The dark green circles represent extended sources within the catalogue and the light green crosses are point sources. Extended sources make up 22$\%$ of the LoBES General and LG-Extended Wide-band catalog. }\label{fig:extpt}
\end{center}
\end{figure}

\subsection{Classification of sources}
Following the method described in \citet{Franzen:2015, Franzen:2019}, we identify extended sources within our catalogue using the ratio of the measured integrated flux density, $S_{\mathrm{int}}$, to the peak intensity, $S_{\mathrm{peak}}$, for each source. To detect source extension at the 2$\sigma$ level
\begin{equation}
\ln \left(\frac{S_{\mathrm{int}}}{S_{\mathrm{peak}}}\right) > 2 \sqrt{\left(\frac{\sigma_{\mathrm{int}}}{S_{\mathrm{int}}}\right)^2 + \left(\frac{\sigma_{\mathrm{peak}}}{S_{\mathrm{peak}}}\right)^2}.
\end{equation}
Here $\sigma_{\mathrm{peak}}$ and $\sigma_{\mathrm{int}}$ are the uncertainties in the measured peak and integrated flux densities, respectively. We take these to be the sum quadrature of the Gaussian fitting uncertainties reported by \textsc{pybdsf} and an overall flux density scale uncertainty of 5$\%$, calculated in Section \ref{sec:err}. 
Figure \ref{fig:extpt} shows the ratio of the integrated to peak intensity densities as a function of the signal-to-noise ratio of the source for detected sources in the LoBES wide-band image at 200~MHz (with resolution of 77~arcsec). Extended sources in this figure are represented by dark green circles; these sources make up 22$\%$ of sources in the Wide-band catalogues. 

\subsection{Source counts}
Here we investigate the source counts of the radio sources within the LoBES EoR0 fields. Source counts as a function of flux density are important for understanding radio source populations and can provide more realistic estimates of the expected foreground contamination in simulations for EoR experiments \citep[e.g.][]{Murray:2017, Nasirudin:2020}. Source counts can also provide an objective way of comparing the results from different surveys. 

\begin{figure*}[t!]
\begin{center}
\includegraphics[scale=0.4]{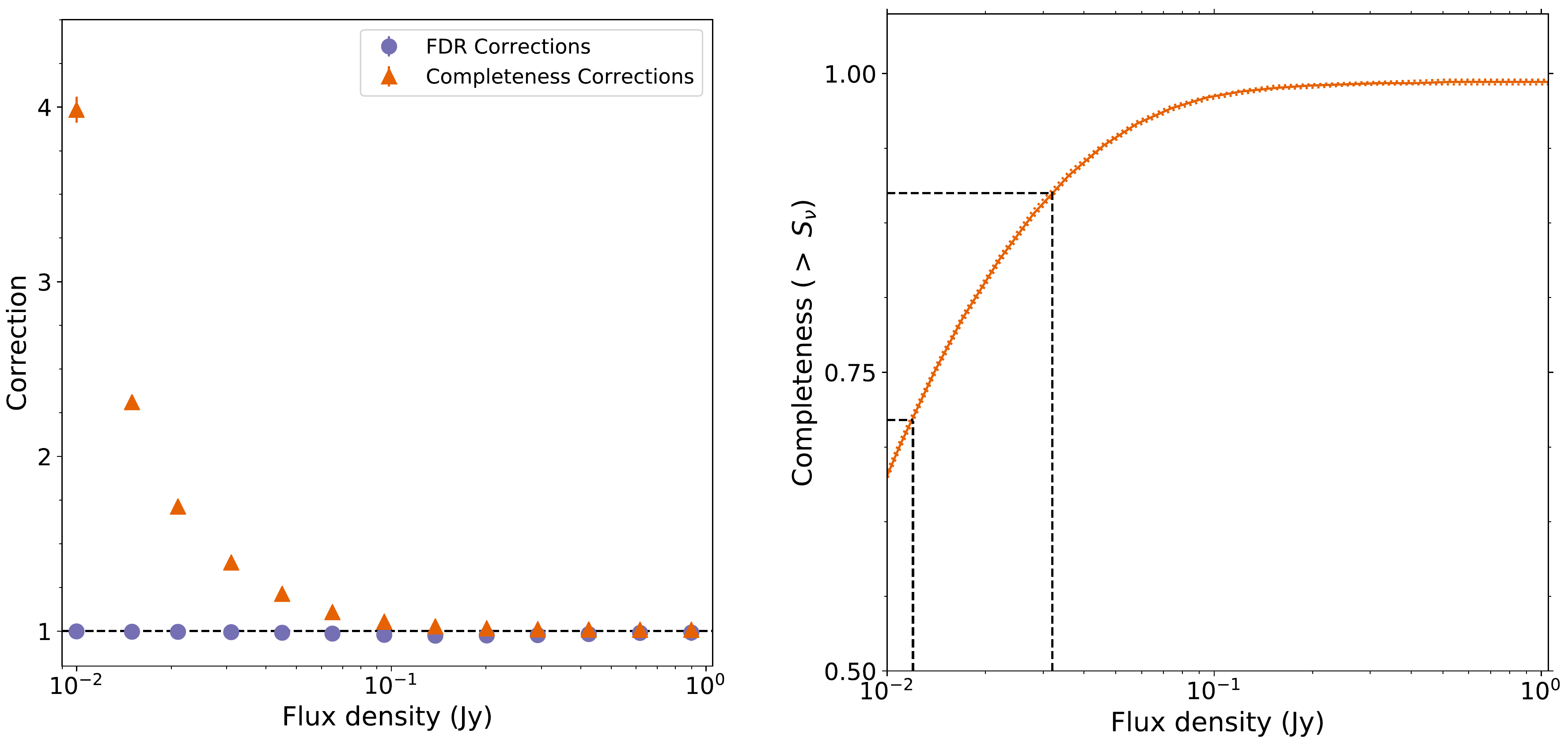}
\caption{The left-hand panel shows the false detection rate (purple circles) and the completeness correction (orange triangles) calculated for the LoBE survey in 15 flux density bins. The correction factors are dominated by the completeness correction, especially at low flux densities. The completeness of the survey is shown on the right. The dashed lines indicate the completeness at 10.5~mJy (70$\%$ complete), the minimum detection threshold of the survey, and the 90$\%$ completeness level at 32~mJy. }\label{fig:srccorr}
\end{center}
\end{figure*}

The chance of detecting sources of various sizes and flux densities in our survey is a function of their location within the survey images as well as their signal-to-noise ratio. This means that the source counts from using \textsc{pybdsf} will not reflect the true extragalactic source count distribution. Several correction factors need to be calculated and applied to the source counts generated from our source finding procedure in order to calculate the expected extragalactic source counts. We outline these correction factors in the following.

\subsubsection{False detection rate}\label{sec:fdr}

To account for noise spikes and artefacts in the images that \textsc{pybdsf} identified as real sources, we use the false detection rate (FDR). From the symmetry of the image noise, we expect that positive noise spikes will have counterpart negative spikes. We will be able to detect these negative spikes in the inverse (negative) image. Using the inverted wide-band images, we run \textsc{pybdsf} using the same parameters as when creating the source catalogue. However, this over estimates the FDR at high flux densities, as \textsc{pybdsf} uses regions of high rms around bright sources to avoid finding and fitting image artefacts around such objects \citep{Hale:2019}.  In the inverted image there are no negative counter-parts to the bright sources in the image, thus it is more likely that negative peaks near bright sources may be counted as sources in the inverted image while their positive counter-parts were not detected.  Additionally, the FDR should generally be an issue at faint flux densities where noise peaks may be confused with sources. To correct the FDR we exclude regions around bright sources by masking out circular regions of 5~arcminutes around sources with flux densities greater than 5~Jy. 

To estimate the FDR we first bin the sources we originally detected in our wide-band image by flux density. We then calculate the number of inverted-image sources that are detected in each of these bins. The fraction of real sources in the $i$th flux density bin is given by
\begin{equation}
f_{\mathrm{real},i} = \frac{N_{\mathrm{catalogue},i} - N_{\mathrm{inverted},i}}{N_{\mathrm{catalogue},i}}
\end{equation}
This fraction is applied multiplicatively to the raw source counts in each flux density bin for the wide-band catalogue and the uncertainty in the FDR is calculated using Poisson errors in the individual number counts. We show the FDR correction factors we used to correct our source counts in the left-hand panel of Figure \ref{fig:srccorr} (indicated by the purple circles) and list the factors for each flux density bin in Table \ref{table:difcnt}. 

\begin{figure*}[t!]
\begin{center}
\includegraphics[scale=0.6]{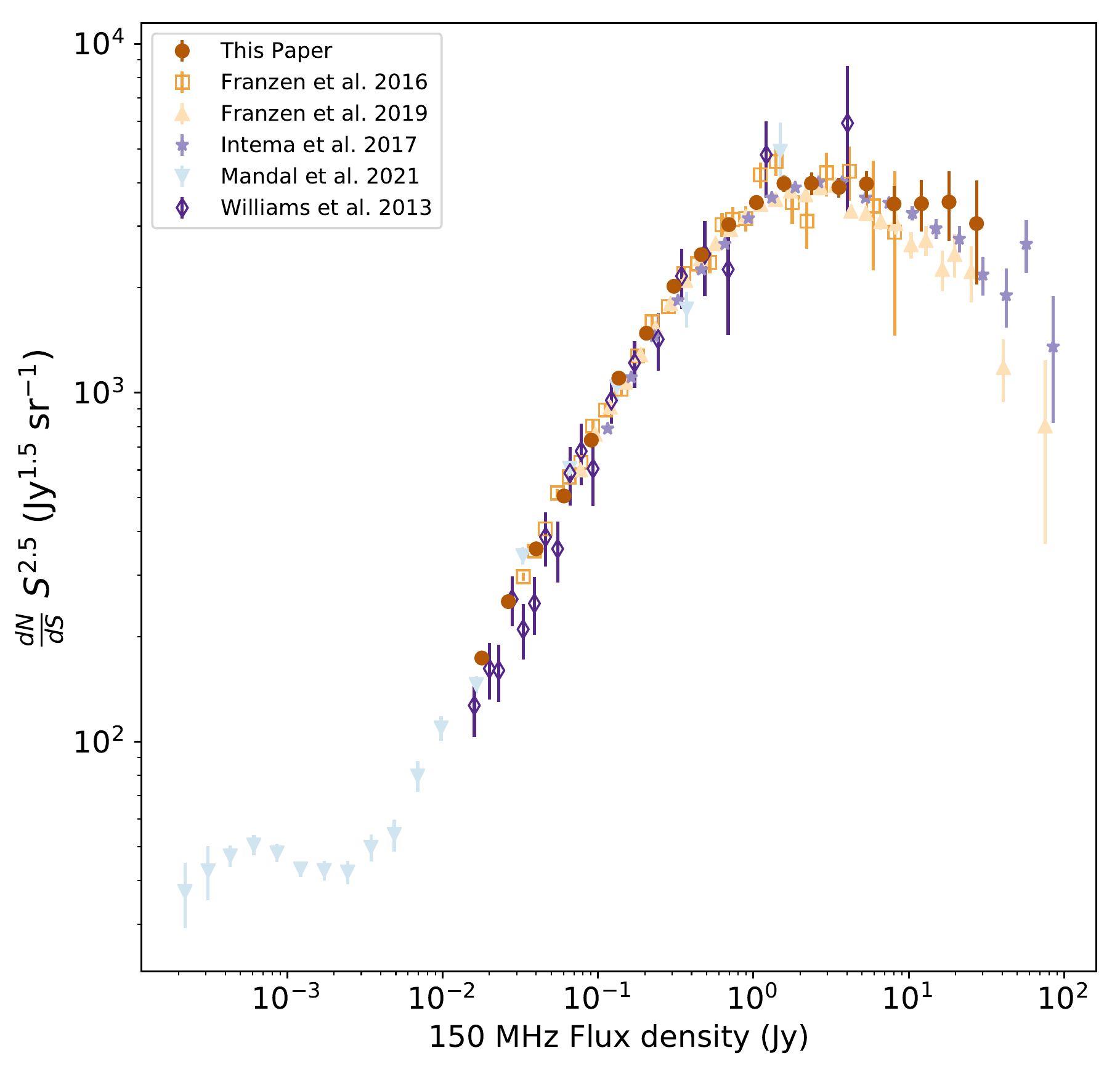}
\caption{Corrected source counts from the LoBE survey (filled brown circles) compared to other survey source counts at $\sim$100~MHz. The comparison surveys included are: MWA results from \citet{Franzen:2016} (orange open squares) and \citet{Franzen:2019} (filled light orange triangles); GMRT results from \citet{Intema:2017} (filled purple stars), and \citet{Williams:2013} (dark purple open diamonds); LOFAR results from \citet{Mandal:2021} (light blue filled inverted triangles)}. The LoBES source counts are in agreement with previous surveys and with the upgrades to the MWA, the sources counts are now becoming competitive with the deepest source counts from the GMRT. \label{fig:srcnt}
\end{center}
\end{figure*}

\subsubsection{Source completeness}
The completeness of a catalogue is the probability that all sources above a given flux density are included in the catalogue. Several factors can have an effect on the completeness of a catalogue. Eddington bias is the result of noise redistributing weak sources into higher flux density bins \citep{Eddington:1913}. This bias is significant near the detection limit of the survey and may boost counts in the faintest bins. Additionally, source finding algorithms, including \textsc{pybdsf}, use peak intensity values to identify sources within an image. This creates a bias against detecting extended sources (termed resolution bias). As the size of a resolved source increases, the ratio of the peak to integrated flux density decreases. Thus an extended source of the same integrated flux density as a point source could have a peak intensity (in units of Jy beam$^{-1}$) that is undetectable by peak-flux-finding algorithms.  Lastly, the variation in the image sensitivity for the sky area covered by the LoBE survey means that faint sources are not detectable across the full survey region -- this is taken into account by considering the visibility area for sources in different flux density bins. 

Following the method outlined by \citet{Williams:2016, Hale:2019, Franzen:2019}, we quantify the various biases in our catalogue by injecting 33000 simulated sources into our set of five wide-band images. These sources are randomly distributed throughout the survey area but with an enforced minimum distance of 5~arcminutes between any two simulated sources. We do not limit the location of a simulated source with respect to the actual sources in the image -- this allows us to account for source confusion. We also make a random 10$\%$ of these sources extended (i.e. source sizes larger than the image beam size). It is important to generate realistic flux densities for these sources and extend the simulated flux densities to well below the 5$\sigma$ detection limit in order to accurately account for Eddington Bias. \citet{Intema:2011} and \citet{Williams:2013} showed that deep source counts at 153~MHz are well represented by d$N$/d$S$ $\propto$ $S^{-1.59}$. The flux densities of the simulated sources were then drawn randomly from this power-law distribution between 5~mJy and 10~Jy and then scaled to 200~MHz using a spectral index of $-$0.8. To generate good statistics we create 100 of these simulations. 

To inject the sources into the images we used \textsc{aeres} from the \textsc{aegean} package. For each simulation we then performed the same source finding procedures used to generate the LoBES wide-band catalogues. The only step excluded in this procedure was the final visual inspection of sources to identify image artefacts and components of large extended sources. We expect the components of the large extended sources will be detected in the simulations and original LoBES wide-band images in the same manner. Additionally, the FDR and masking outlined in Section \ref{sec:fdr} should account for image artefacts. Thus, in our final source count analysis we use the LoBES wide-band catalogue before removing sources identified via visual inspection. 

\begin{table*}
\caption{The Euclidean-normalised differential source counts for the LoBES EoR0 fields scaled to 154~MHz. The range of the flux density bin is given by $S_{\textrm{range}}$, with a centre flux density of $S_{\textrm{c}}$. $N$ is the total number of uncorrected sources per bin and the last column gives the corrected normalised source counts ($S^{2.5}$ d$N$/d$S$).}
\begin{tabular}{cllccl}
\hline
$S_{\textrm{range}}$ & $S_\textrm{c}$ & $N$ & FDR & Completeness Correction & Corrected $S^{2.5}$ d$N$/d$S$ \\
(Jy) & (Jy) & & & & (Jy$^{3/2}$ sr$^{-1}$) \\
\hline
0.015 -- 0.021 & 0.018 & 7620  & 0.99$\pm$0.02 & 3.29$\pm$0.09   & 174$\pm$6 \\
0.021 -- 0.032 & 0.027 & 12682 & 0.99$\pm$0.01 & 1.94$\pm$0.03   & 252$\pm$6 \\
0.032 -- 0.048 & 0.040 & 12501 & 0.99$\pm$0.01 & 1.43$\pm$0.02   & 357$\pm$8 \\
0.048 -- 0.072 & 0.060 & 11732 & 0.99$\pm$0.01 & 1.20$\pm$0.01   & 506$\pm$10 \\
0.072 -- 0.108 & 0.090 & 9959  & 0.98$\pm$0.01 & 1.080$\pm$0.008 & 731$\pm$14 \\
0.108 -- 0.164 & 0.136 & 8396  & 0.98$\pm$0.01 & 1.080$\pm$0.008 & 1101$\pm$21 \\
0.164 -- 0.247 & 0.205 & 6359  & 0.98$\pm$0.02 & 1.034$\pm$0.007 & 1481$\pm$32 \\
0.247 -- 0.370 & 0.308 & 4767  & 0.97$\pm$0.02 & 1.016$\pm$0.004 & 2019$\pm$50 \\
0.370 -- 0.556 & 0.463 & 3215  & 0.98$\pm$0.02 & 1.010$\pm$0.005 & 2488$\pm$76 \\
0.556 -- 0.836 & 0.696 & 2107  & 0.99$\pm$0.03 & 1.008$\pm$0.005 & 3033$\pm$118 \\
0.836 -- 1.257 & 1.047 & 1325  & 0.99$\pm$0.04 & 1.008$\pm$0.005 & 3511$\pm$176 \\
1.257 -- 1.890 & 1.573 & 812   & 0.99$\pm$0.05 & 1.008$\pm$0.005 & 3979$\pm$216 \\
1.890 -- 2.841 & 2.365 & 438   & 0.99$\pm$0.06 & 1.007$\pm$0.006        & 3983$\pm$306 \\
2.841 -- 4.271 & 3.556 & 231   & --------      & --------        & 3877$\pm$248 \\
4.271 -- 6.420 & 5.346 & 128   & --------      & --------        & 3970$\pm$340 \\
6.420 -- 9.651 & 8.036 & 61   & --------       & --------        & 3476$\pm$431 \\
9.651 -- 14.51 & 12.08 & 33    & --------      & --------        & 3483$\pm$585 \\
14.51 -- 21.81 & 18.16 & 18    & --------      & --------        & 3521$\pm$799 \\
21.81 -- 32.80 & 27.30 & 9    & --------       & --------        & 3055$\pm$1010 \\
32.80 -- 49.30 & 41.05 & 1    & --------       & --------        & 704$\pm$658 \\

\hline
\end{tabular}\label{table:difcnt}
\end{table*}

From this process we have a set of 100 source catalogues containing both the detected injected sources and original image sources. To calculate the completeness correction, we then use \textsc{stilts} to do a simple cross-match between each of the 100 catalogues with the original injected source catalogue, matching sources within the resolution of the wide-band image. This identifies which of the injected sources are recovered in each of the 100 simulations. To account for source confusion between the injected sources and the original LoBES sources, we then cross-match the identified sources with the original LoBES wide-band catalogue. We remove matched sources that have measured source positions closer to a matched LoBES source than to the matched injected source. 

The correction factor in the $i$th flux density bin is then calculated as
\begin{equation}
\textrm{Completeness Correction}_i = \frac{N_{\mathrm{injected},i}}{N_{\mathrm{recovered},i}}
\end{equation}
The final correction factors in each of the flux density bins is taken as the median of the correction values calculated in the 100 simulations; the uncertainties are taken as the 16 and 84th percentile values. The left-hand panel of Figure \ref{fig:srccorr} shows the completeness correction factors for the LoBES catalogue (orange triangles); the values for each bin are also listed in Table \ref{table:difcnt}. In this figure (and in the table) it is evident that the completeness correction contributes the most to the overall correction of the source counts, with the largest correction factors occurring in the lowest flux density bins.  The completeness of the catalogue at a given flux density is then determined by integrating the detected fraction of sources (i.e. the inverse of the completeness correction value) upwards from a given flux density limit. The completeness of the LoBES catalogue from 10~mJy to 1~Jy is shown in the right-hand panel of Figure \ref{fig:srccorr}. We estimate that the catalogue is 90$\%$ complete above a flux density of 32~mJy.

\subsubsection{Final source counts}
To calculate the corrected source counts for LoBES we combine the FDR and completeness corrections multiplicatively, and multiply the source counts in each flux density bin by the appropriate factor. The uncertainties in the FDR, completeness correction, and source counts are combined in quadrature. To compare to other surveys we scale the flux density bins to 154~MHz assuming a spectral index of -0.8. The corrected source counts bewteen 15~mJy and 50~Jy are show in Figure \ref{fig:srcnt}, with the values in each scaled flux density bin listed in Table \ref{table:difcnt}. 

Figure \ref{fig:srcnt} compares the source counts measured here (brown circles) to those in the literature, including a deep survey of the Bo\"otes region at 150~MHz using the GMRT (dark purple open diamonds \citet{Williams:2013}), the source counts estimated from the TGSS ADR1 survey (purple stars, \citet{Intema:2017}), source counts estimated using a deep MWA phase I images at 154~MHz of the EoR0 region (orange squares, \citet{Franzen:2016}), 154~MHz source counts from the MWA GLEAM survey (light orange triangles, \citet{Franzen:2019}), and 150~MHz source counts from the LOFAR Two Meter Sky Survey (LoTTS) Deep Fields (light blue inverted-triangles \citet{Mandal:2021}). Note that this list of papers is not exhaustive. The LoBES source counts are in good agreement with the literature, and the deepest LoBES source counts are comparable to the deepest source counts from the GMRT surveys at 154~MHz. 

\section{Impact of improved model}
 To investigate the performance of the new LoBES sky model as compared to the current Australian MWA EoR sky model, we use each sky model in one of the standard MWA EoR collaboration pipelines (as outlined in \citet{Jacobs:2016}) for estimating the EoR power-spectrum. We then compare the data residuals resulting from each run of the pipeline. We use real data from Phase I of the MWA in this investigation. We choose to use MWA Phase I data to test the sky models because the MWA Phase I configuration is sensitive to the spatial scales needed to do EoR science.
 
 The MWA EoR pipeline we use involves first calibrating and removing foreground sources from the MWA data using the RTS and then measuring the power-spectrum of the residuals using the Cosmological H I Power Spectrum Estimator (CHIPS; \citet{Trott:2016a}). Because it is not currently computationally feasible to use a horizon-to-horizon sky model for either calibration or source removal using the MWA EoR pipelines, only a subset of the total sky is used to complete these steps \citep{Jacobs:2016}. Previous MWA EoR limits using the RTS and CHIPS pipeline select the 1000 apparently brightest sources for each snapshot observation to use during calibration and source removal \citep{Yoshiura:2021, Trott:2020}. Similarly in the following analysis we use the primary beam model from \citet{Sokolowski:2017} to choose the apparently brightest 1000 sources from the LoBES sky model for each observation. To be consistent, we identify the source models associated with these 1000 sources in the current Australian MWA EoR sky model for each observation to use in our comparison. The source models from either LoBES or the current Australian MWA EoR sky model for these 1000 sources are used for both calibration and source removal.
 
To assess the relative impact of the new sky model, we use the two-dimensional power-spectrum (2D PS). The 2D PS is formed by averaging $k$-modes associated with angular scales on the sky, and plotting the power as a function of angular $k$-modes (modes perpendicular to the line of sight; $k_{\perp}$), and spectral $k$-modes (modes parallel to the line of sight; $k_{\parallel}$). Generally, it is expected that astrophysical foregrounds with be spectrally smooth and contaminate only small values of $k_{\parallel}$. However, the inherent chromaticity of the radio interferometer will cause the foreground emission to spread to higher values of $k_{\parallel}$ where measurements of the EoR are expected to occur. This creates the so-called `foreground wedge' and the remainder of the 2D PS parameter space is dubbed the `window' \citep{Datta:2010, Trott:2012, Thyagarajan:2013, Vedantham:2012}. Additionally, frequency-dependent calibration errors will couple with residual foreground power and further leak power into the EoR window \citep[e.g.][]{Barry:2016, Trott:2016b, Offringa:2016, Ewall-Wice:2017}. Because the power of EoR signal decreases with increasing $\|k\|$, the most sensitive measurements are expected to be in the lower-left corner of the EoR window. Thus it is critical to reduce the amount of foreground power that leaks from the wedge to the window.  

\subsection{Observations and reduction}\label{sec:data}
To test our sky model using MWA Phase I data,  we select sixteen zenith scans (encompassing roughly 30 minutes of data) from the September, 01 2014 observing campaign of the MWA EoR0 field, observed in the 170.2~--~199.7~MHz frequency band. These data were chosen based on the ionospheric quality metrics from \citet{Jordan:2017}, choosing data associated with quiet ionospheric conditions. 

Using the RTS we process each observation scan separately. During calibration, a compound calibrator model is generated consisting of the 1000 apparently brightest sources within the observation field of view. These sources are combined into a single calibrator in order to achieve a high signal-to-noise ratio. The direction-independent (DI) Jones matrices for each MWA tile are then computed by fitting the uncalibrated visibilities with the compound calibrator model. To create this compound calibrator model we first selected the 1000 apparently brightest sources from the LoBES catalogue using the MWA primary beam model from \citet{Sokolowski:2017}. The associated sky models for these 1000 sources from the current Australian MWA EoR sky model are then identified. Initially we used these two sets of sky models for the same 1000 sources to carry out two independent runs of the RTS to do DI calibration of the MWA test data. However, we found that using the same 1000 sources from each sky model to perform this calibration step caused differences in the overall flux density scale between the two sets of calibrated visibilities. This is due to differences in the source flux densities between the current Australian MWA EoR sky model and LoBES for the same 1000 sources. To ensure a fair comparison between the two sky models, we use the sky models for the 1000 apparently brightest sources from LoBES to perform DI calibration for both runs of the MWA EoR pipeline.

Using the same 1000 sources used for DI calibration, we then run the RTS a second time to perform direction-dependent (DD) calibration. During DD calibration for the two runs of the MWA EoR pipeline we used either the source models from the LoBES catalogue or the source models from the current Australian MWA EoR sky model for these 1000 sources. The DD calibration in the RTS is based upon the `peeling' technique \citep{noordam04}. Using the DI calibration solution and the source models for the 1000 calibrator sources, all 1000 sources are directly subtracted from the observed visibilities. The sources are then ranked based on their apparent brightness. For each source in the ranked source list, starting with the top ranked source, the RTS then adds back the calibrator source and phase-rotates the visibilities to the location of this source. The source position is then corrected for ionospheric refraction in the source direction by fitting a phase ramp to the phased visibilities. For only the top five brightest sources in the ranked list the full antenna-based DD gain solutions are calculated and applied. This DD calibration process essentially peels the five brightest sources and directly subtracts the remaining 995 sources. However for simplicity we will use the word `peeling' when referring to any source subtraction performed. For more details on this process see \citet{Mitchell:2008}.

For each observation we find about 38$\%$ of the peeled sources are modelled by a single point source in both catalogues; roughly 58$\%$ of the sources modelled by a point source in the current Australian MWA EoR sky model are modelled by at least one Gaussian in the LoBES catalogue; and finally about 4$\%$ of the peeled sources are modelled by at least one Gaussian in both sky models, but for all these sources LoBES uses more Gaussian components to model each source than the current sky model. This means that overall 62$\%$ of the peeled sources are modelled by more components in LoBES than in the current sky model. 

\begin{figure}[t!]
\centering
\begin{subfigure}[b]{\columnwidth}
\centering
\includegraphics[width=\columnwidth]{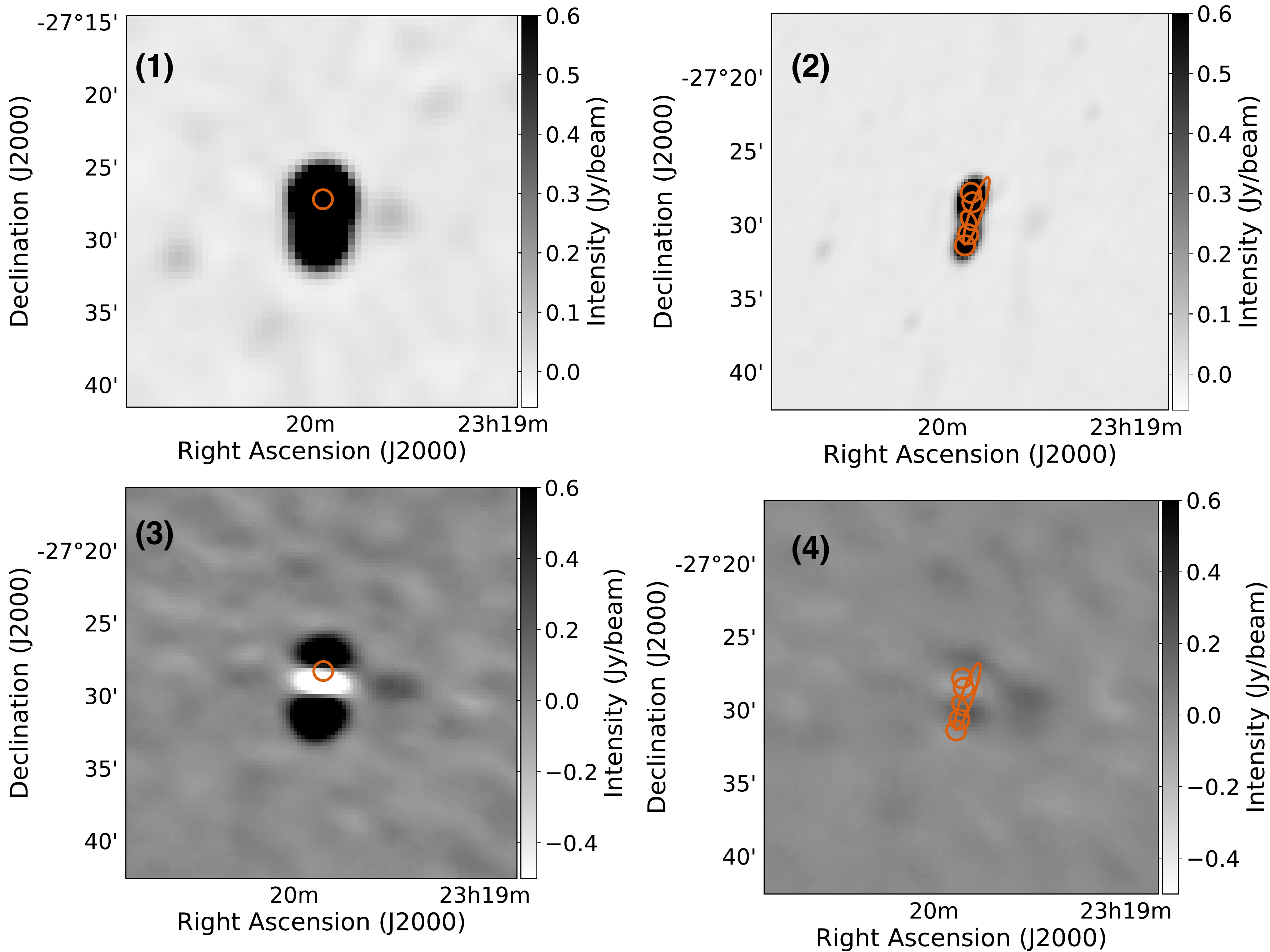}
\caption{PKS 2317-27}
\vspace{5mm}
\end{subfigure}

\begin{subfigure}[b]{\columnwidth}
\centering
\includegraphics[width=\columnwidth]{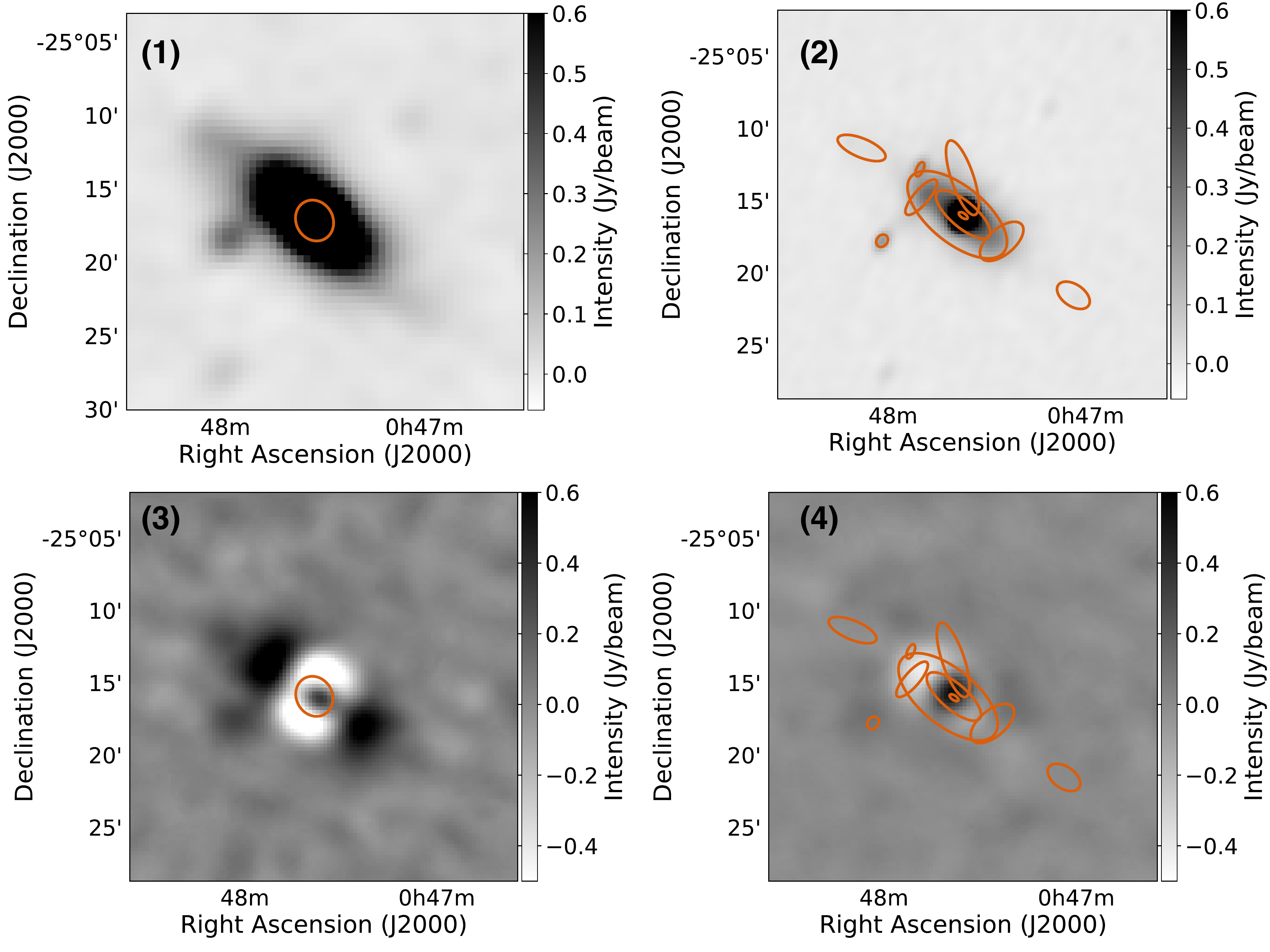}
\caption{NGC 253}
\end{subfigure}
\caption{Comparison between the current Australian MWA EoR sky model and the LoBES sky model for two sources located with the EoR0 field. Shown are the GLEAM images for both sources  in panels (1);  the LoBES wide-band image in panels (2); panels (3) show the peel residuals for the current Australian MWA EoR Sky Model using our set of 2014 MWA Phase I test data; and panels (4) show the peel residuals in the 2014 MWA Phase I test data using the LoBES Sky Model. The current Australian MWA EoR sky model for each source is overlaid in the orange ellipses in panels (1) and (3); similarly the LoBES Sky Model is overlaid in panels (2) and (4). Comparing the peel residuals for the two models it is evident that the LoBES source model removes more emission than the current sky model for these two sources.  }\label{fig:peelcomp} 
\end{figure}

Using \textsc{wsclean} we imaged both sets of calibrated and peeled observations, integrating over the full 30 minutes and 30.72~MHz bandwidth; we used the same clean parameters listed in Section \ref{sec:imaging} for this imaging. Comparing the residuals for the removed sources in these two images by eye, it is clear that there is significant improvement for some of the more complex sources in the field. Figure \ref{fig:peelcomp} shows two examples of improved residuals using the LoBES sky model for two of the brightest extended sources in the EoR0 field. 

\begin{figure}[t!]
\begin{center}
\includegraphics[width=\columnwidth]{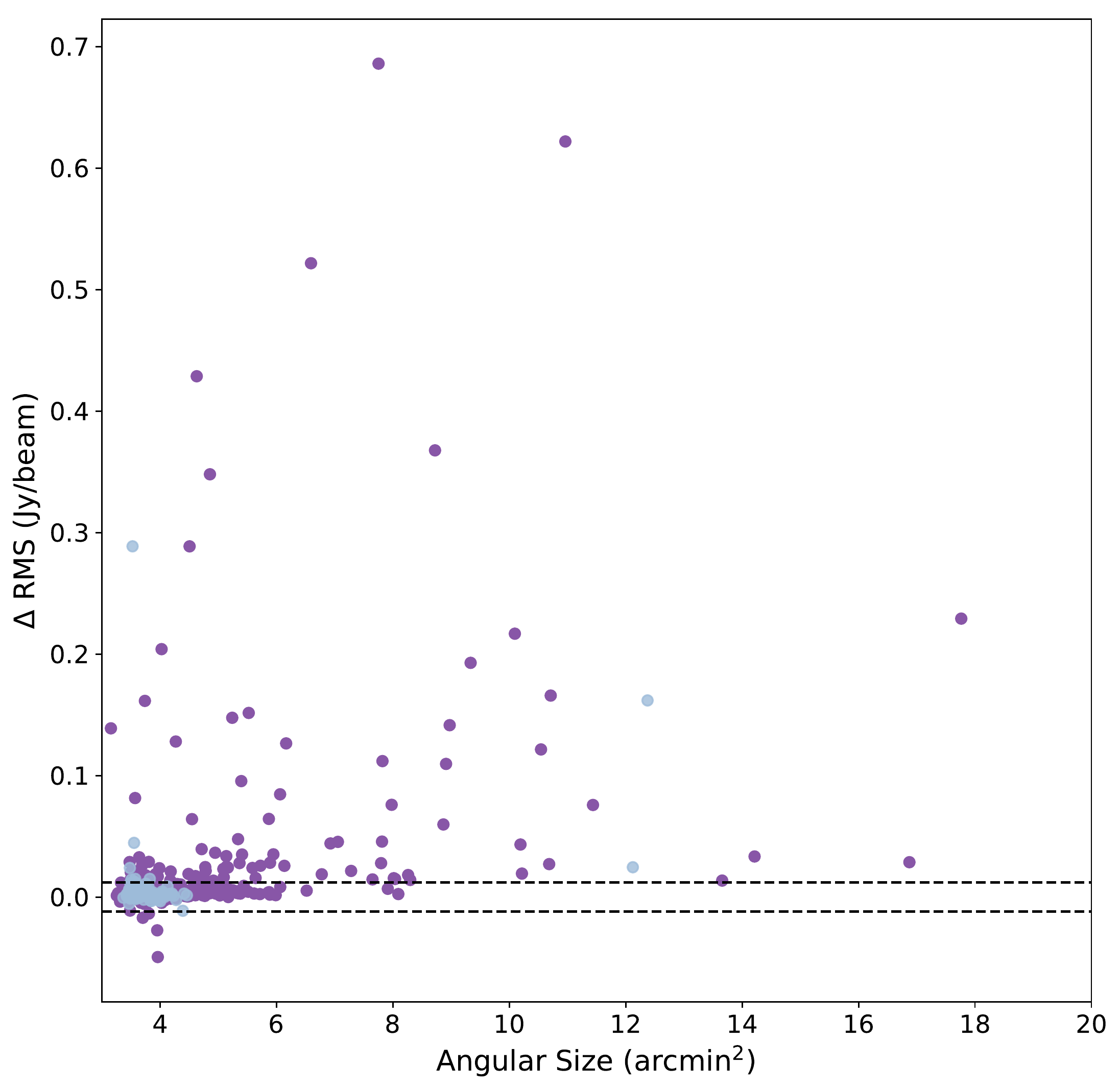}
\caption{Differences in residual rms between the current Australian MWA EoR sky model and the LoBES sky model as a function of angular size of removed sources in our 2014 test data set. Purple circles indicate sources that are fit with more components in LoBES than in the current sky model; blue circles represent sources fit with the same number of components in the two sky models. Dashed lines indicate 1$\sigma$ rms values in the integrated images. Sources with positive differences greater than the image noise have smaller residuals using the LoBES models (27$\%$ of the sources). Note that the largest sources peeled from the data have improved rms value when using LoBES models. }\label{fig:diffs}
\end{center}
\end{figure}

\begin{figure*}[t!]
\begin{center}
\includegraphics[scale=0.48]{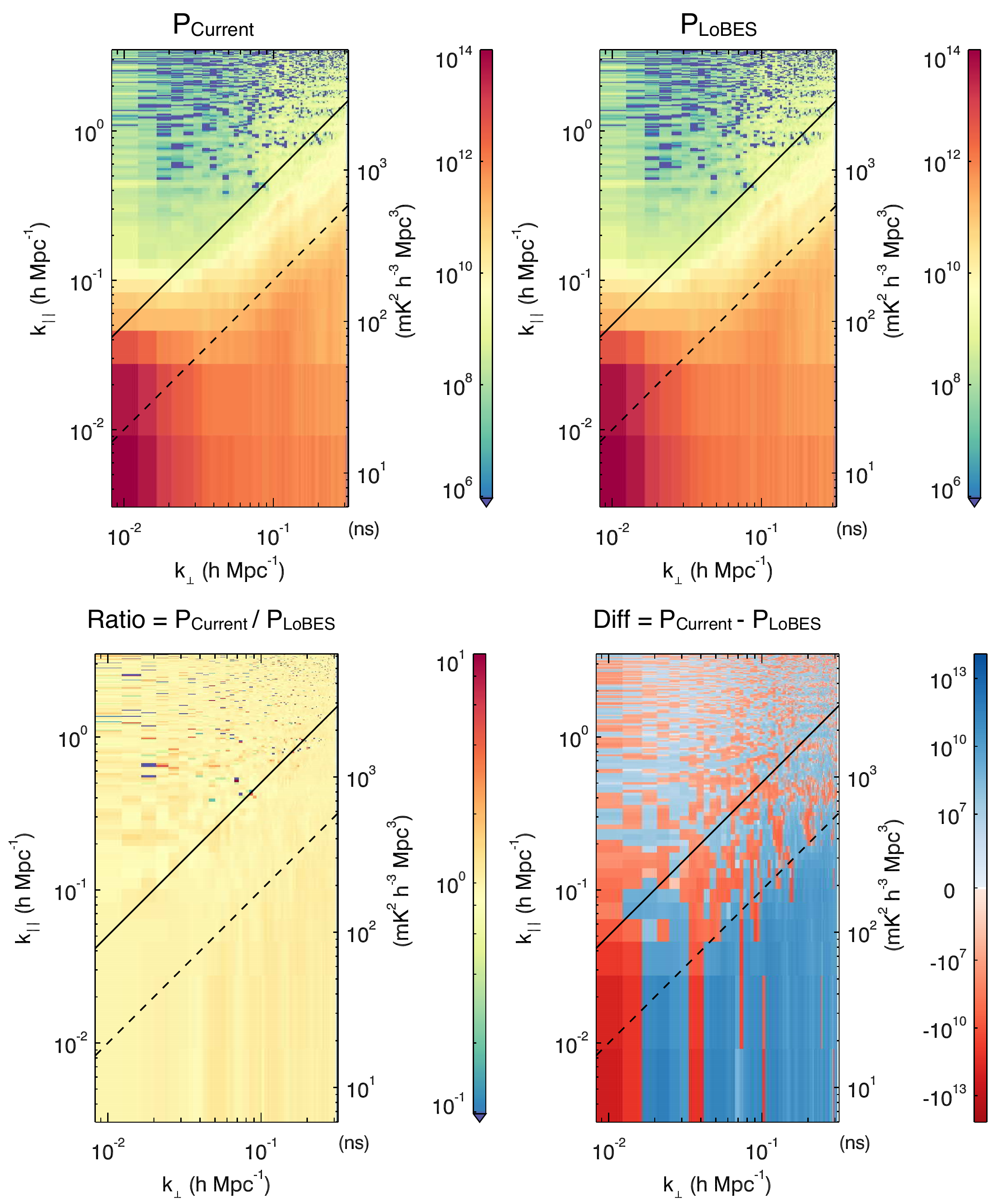}
\caption{The resulting 2D PS from using the current Australian MWA EoR sky model (top left) and LoBES sky model (top right) to peel sources from 30 minutes of MWA Phase I data. The bottom row shows the ratio (left) of the residual power from the current model to the LoBES model, and difference (right) in residual power between the current model and LoBES. Also indicated are the primary field of view and horizon lines. These indicate the expected contamination areas for sources in the primary field of view and sidelobes. Within the difference plot its clear that the LoBES catalogue better removes foreground power on small angular scales within the foreground wedge.}\label{fig:2dps_data}
\end{center}
\end{figure*}

To quantify the differences in the source residuals we measure the rms within regions centred on the peeled sources in each of the integrated images. In this analysis we assume that smaller rms values are indicative of a better source subtraction. Figure \ref{fig:diffs} shows the differences in the residual rms as a function of the convolved angular source size as measured in LoBES. The dark purple points indicate sources that are fitted with more components in the LoBES sky model than their counterpart in the current Australian MWA EoR sky model; the light blue points are for sources with the same number of fit components in both sky models. In this figure, large positive differences in the residual rms indicate that the LoBES sky models reduced the local rms significantly as compared to the current Australian MWA EoR sky model.  Roughly 88~$\%$ of the sources with a different number of model components have a smaller rms using the LoBES model, however only about 27~$\%$ of these sources have a rms difference that is significant, meaning that the difference is greater than the expected 1$\sigma$ image noise (indicated by the dashed line in Figure \ref{fig:diffs}). The improvement for the sources fit with the same number of components, which is dominated by sources fit with a single component, is marginal with only 5$\%$ of these sources showing a significant improvement in the residual rms for the LoBES model. Figure \ref{fig:diffs} shows that the multi-gaussian source modelling had the most significant impact on the largest peeled sources; as the source size increases the proportion of sources with significant improvement in the residual rms increases. This is consistent with the larger differences being found in the sources with more model components in LoBES as these tend to be the largest sources within the catalogue.


We also investigate how the differences between these two sky models impact the 2D PS. The top two panels of Figure \ref{fig:2dps_data} show the resulting 2D PS for the current Australian MWA EoR sky model (top left) and LoBES sky model (top right) in the YY-polarisation. To more easily see any differences between the 2D PS for the two sky models we also include the ratio plot for the current sky model residual power to the LoBES model residual power (bottom left) and the difference plot, where the values represent the current sky model minus the LoBES model (bottom right). In each of the 2D PS we also indicate the `primary field of view' line (dashed) and the `horizon' line (solid) which indicate the expected contamination limits for sources in the primary field of view and the sidelobes. The blue areas within the difference plot show regions of the 2D PS where the residual power is higher for the current Australian MWA EoR sky model as compared to the LoBES sky model. In Figure \ref{fig:2dps_data}, the consistently blue region restricted to sources within the primary field of view and on small angular scales (large $k_{\perp}$) is statistically significant as compared to the residual power within this region of the LoBES 2D PS. This blue region indicates that the LoBES sky model performed better at modelling emission on small angular scales. This improvement is consistent with the difference in resolution for the two catalogues, where the LoBES catalogue includes higher resolution modelling and is expected to remove smaller scale emission better. 

There is no significant difference between the two catalogues on larger angular scales. We believe that any possible differences in the residual power between the two models is being overwhelmed by other systematics in the data. One obvious systematic that is expected on these angular scales is the contribution of foreground power from the Milky Way. The EoR0 field is known to contain significant diffuse polarised emission from the Galactic plane \citep{Lenc:2017, Bernardi:2013}. In our analysis here we do not consider the impact of excluding this foreground in our sky models and it is likely this component dominates residual power on large angular scales. Additionally there may be other data quality metrics, beyond ionospheric quality, that need to be considered. Both \citet{Barry:2019} and \citet{Li:2019} have shown that improvements in the PS generation pipeline and careful selection of data can have a significant impact on the residual power in the PS. For example \citet{Barry:2019} showed that removing data that contain low-level RFI, missed by the normal flagging algorithms \citep{Wilensky:2019}, yielded improved PS limits. Preliminary investigations into what could be the limiting systematic, revealed a variation in the quality of the source peeling from observation to observation for sources located towards the edge of the field-of-view. We used the analytic beam model to approximate the MWA primary beam when calibrating and peeling using the RTS. This beam model is known to be inaccurate far from the phase centre \citep{Sokolowski:2017, Sutinjo:2015}. Future investigations into the peel quality will determine whether using improved beam models will remove this variation in the peel residuals.

\section{Summary}
In this paper we present the first results from the LoBE survey. The main goal of this survey is to use the new extended configuration of the MWA to provide higher resolution, multi-frequency modelling of the foreground sources within the main MWA EoR observing fields and their eight neighbouring fields. Here we focus on the MWA EoR0 field, centred at 0.0 hours and $-$27.0~degrees, and its four neighbouring fields. The survey covers the frequency range of 100 -- 230~MHz with 16 spectral measurements, and reaches an average sensitivity of 2.1~mJy~beam$^{-1}$ in a deep wide-band (60~MHz) image. This half of the survey covers an area of 3069~degrees$^2$ and we identify 80824 sources, including 45 large-extended sources, within this sky region. Given the wide frequency coverage of our survey we perform spectral modelling for 78$\%$ of the sources in our catalogue, taking into account any spectral curvature that is observed. The derived spectral indices for sources best fit with a standard power-law agree with those previously measured in the literature. 

We calculate that our source catalogue is 70$\%$ complete at 10.5~mJy and 90$\%$ at 32~mJy. The differential source counts measured in the LoBE survey, after correcting for various bias factors, are in good agreement with other surveys at $\sim$100~MHz and similar sensitivities. While still not reaching the sensitivities of LOFAR, the upgrades to the MWA have resulted in greater imaging capabilities, reaching sensitivities comparable to the deepest surveys from the GMRT at similar frequencies. Thus, we expect future all-sky Southern Hemisphere surveys using the MWA phase II capabilities, like GLEAM-X (Hurley-Walker in prep.), will be able to provide complementary source counts and catalogues to those using Northern Hemisphere low-frequency telescopes. This will be particularly important for the future SKA Low, as sky models developed from MWA surveys will be crucial to initially calibrate data from this telescope. 

Testing the new LoBES models using MWA Phase I data, we find an improvement in the peel residuals when using the new models as compared to using the current Australian MWA EoR sky model. This is particularly true for the largest sources in the EoR0 field. Comparing the 2D PS of the residual power for the two catalogues, we find that the LoBES sky models remove more foreground emission on small angular scales within the foreground wedge. The residual large scale power for the two catalogues is not significantly different and likely limited by other systematics. We suggest various factors which could be contributing to the overlying systematics in the data, including foreground emission from the Galactic plane. The bright foreground emission from the Milky Way has thus far been ignored in MWA EoR analysis. Once the LoBES extragalactic catalogue is complete (in a forthcoming paper 2), the next major step in foreground modelling for the MWA experiment will be to include models of the diffuse Galactic emission. 


\begin{acknowledgements}
CRL thanks Q. J. Roper for the many helpful discussions about statistics that added to the analysis presented in this paper. CMT is supported by an ARC Future Fellowship under grant FT180100321. This work was partially supported by the Centre for All Sky Astrophysics  in  3 Dimensions (ASTRO3D), an Australian Research Council Centre of Excellence, funded by grant CE170100013.  The International Centre for Radio Astronomy Research (ICRAR) is a Joint Venture of Curtin University and The University of Western Australia, funded by the Western Australian State government. This scientific work makes use of the Murchison Radio-astronomy Observatory, operated by CSIRO. We acknowledge the Wajarri Yamatji people as the traditional owners of the Observatory site. Support for the operation of the MWA is provided by the Australian Government (NCRIS), under a contract to Curtin University administered by Astronomy Australia Limited. We acknowledge the Pawsey Supercomputing Centre, which is supported by the Western Australian and Australian Government. 
\end{acknowledgements}



\bibliographystyle{pasa-mnras}
\bibliography{eor_bib}

\end{document}